\documentclass[conference]{IEEEtran}
\IEEEoverridecommandlockouts
\usepackage{cite}
\usepackage{amsmath,amssymb,amsfonts}
\usepackage{algorithmic}
\usepackage{graphicx}
\usepackage{textcomp}
\usepackage{xcolor}

\usepackage{multirow}
\usepackage{float}
\usepackage{subcaption}
\usepackage{mwe}
\def\BibTeX{{\rm B\kern-.05em{\sc i\kern-.025em b}\kern-.08em
    T\kern-.1667em\lower.7ex\hbox{E}\kern-.125emX}}
\begin{document}

\title{Smart Palm: An IoT Framework for Red Palm Weevil Early Detection\\
\thanks{This  work  is  supported  by  Elm Company, within the Robotics and Internet-of-Things Lab;  also  by  the  Prince Sultan University. We would like also to thank Dr. Mona Alduailej, Assistant Professor in Computational Statistics in Princess Norah University, for her insights in the statistical analysis section. }
}


\author{\IEEEauthorblockN{Anis Koubaa \IEEEauthorrefmark{1} , Abdulrahman Aldawood \IEEEauthorrefmark{3}, Bassel Saeed \IEEEauthorrefmark{1}, \\Abdullatif Hadid \IEEEauthorrefmark{1}, Mohanned Ahmed \IEEEauthorrefmark{1}, Abdulrahman Saad \IEEEauthorrefmark{4}, Hesham Alkhouja \IEEEauthorrefmark{1}, Mohamed Alkanhal \IEEEauthorrefmark{1}}
\IEEEauthorblockA{\IEEEauthorrefmark{1}Robotics and Internet-of-Things Lab, Prince Sultan University, Saudi Arabia \\
\IEEEauthorblockA{\IEEEauthorrefmark{3} King Saud University, Saudi Arabia.}
\IEEEauthorrefmark{1}(akoubaa)@psu.edu.sa,  
}

}

\maketitle 
\thispagestyle{empty}
\pagestyle{empty}

\begin{abstract}
Smart agriculture is an evolving trend in agriculture industry, where sensors are embedded into plants to collect vital data and help in decision making to ensure higher quality of crops and prevent pests, disease, and other possible threats. In Saudi Arabia, growing palms is the most important agricultural activity, and there is an increasing need to leverage smart agriculture technology to improve the production of dates and prevent diseases. One of the most critical diseases of palms if the red palm weevil, which is an insect that causes a lot of damage to palm trees and can devast large areas of palm trees. The most challenging problem is that the effect of the weevil is not visible by humans until the palm reaches an advanced infestation state. For this reason, there is a need to use advanced technology for early detection and prevention of infestation  propagation. 
In this project, we have developed am IoT based smart palm monitoring prototype as a proof-of-concept that (1) allows to monitor palms remotely using smart agriculture sensors, (2) contribute to the early detection of red palm weevil. Users can use web/mobile application to interact with their palm farms and help them in getting early detection of possible infestations. We used Elm company IoT platform to interface between the sensor layer and the user layer. 
In addition, we have collected data using accelerometer sensors and we applied signal processing and statistical techniques to analyze collected data and determine a fingerprint of the infestation.


\end{abstract}


\section{PROJECT MANAGEMENT}
In this report, we present the results of the four-month project sponsored by Elm company, which aims at developing a proof-of-concept of an IoT system that allows users to monitor palm farms remotely and help them in making early detection of the red palm weevil infestation of palms. The project was executed from February 2019 until May 2019. This report was prepared in the period June-July 2019, after the completion of the project.

In this section, we present project management. 

\subsection{R\&D Team}
\begin{itemize}
	\item \textbf{Senior Researchers}
	\begin{itemize}
		\item Dr. Anis Koubaa, Professor, Department of Computer Science, College of Computer and Information Sciences, Prince Sultan University
		\item Dr. Abdelrahman AlDaoud, Professor, Department of Plant Protection, College of Food and Agriculture Sciences, King Saud University
	\end{itemize} 
	\item \textbf{Students}
	\begin{itemize}
		\item Mr. Bassel Saeed, SE Student, Full-Stack Developer, Department of Software Engineering, College of Computer and Information Sciences, Prince Sultan University
		\item Mr. Abdullatif Hadid, SE Student, Sensor Deployment and Data Collection, Department of Software Engineering, College of Computer and Information Sciences, Prince Sultan University
		\item Mr. Mohanned Ahmed, SE Student, Data Analytics, Department of Software Engineering, College of Computer and Information Sciences, Prince Sultan University
		\item Mr. Abdulrahman Saad, SE Student, Data Analytics, Department of Software Engineering, College of Computer and Information Sciences, Prince Sultan University
		\item Mr. Hesham Alkhouja, SE Student, Experimental Works, Department of Software Engineering, College of Computer and Information Sciences, Prince Sultan University
	\end{itemize} 
\end{itemize} 

\subsection{Research and Development Methodology}
The project is organized in four phases:
\begin{itemize}
	\item \textbf{WP1. Preliminaries} {\textit{Status: Completed 100\%}}
	\begin{itemize}
		\item T1.1. Literature review on smart agriculture solutions
		\item T1.2. Literature review on weevil detection system
	\end{itemize}
	
	\item \textbf{WP2: Sensor Development} {\textit{Status: Completed 100\%}}
	\begin{itemize}
		\item T2.1. Purchase of Arduino sensors kit
		\item T2.2. Development of red palm weevil detector sensor
		\item T2.3. Development of smart palm sensor kit (climate conditions)
	\end{itemize}
	
	\item \textbf{WP3: IoT Data Collection} {\textit{Status: Completed 100\%}}
	\begin{itemize}
		\item T3.1. Development of communication/networking protocols for data collection
		\item T3.2. Integration of smart agriculture sensors with Elm IoT platform
		\item T3.3. Data storage on Elm IoT platform
		\item T3.4. Tracking battery consumption 
	\end{itemize}
	
	\item \textbf{WP4: User Interfaces} {\textit{Status: Completed 100\%}}
	\begin{itemize}
		\item T4.1. Development of mobile/web app for monitoring
		\item T4.2. Development and web dashboards
	\end{itemize}
	
	\item \textbf{WP5: Experimental Integration} {\textit{Status: Completed 100\%}}
	\begin{itemize}
		\item T5.1. System integration and testing
		\item T5.2. Experimental data analysis
		\item T5.3. Validation of infection detection
	\end{itemize}
	
\end{itemize}

In addition to the aforementioned work packages specified in the project proposal, we have also worked an extra tasks, requested by Elm company, related to signal processing techniques for weevil detection. This additional task was executed during June and July after the exams and during the vacation. 
We have established an experimental testbed for data collection and analytics of weevil insect to determine its signature using various signal processing and statistical techniques. 

In this report, we present all the findings and contributions of this project.

\section{RELATED WORKS}

Date palms are the most cultivated trees and contribute significantly to the economy of the Arabian Peninsula, especially Saudi Arabia. There are over 28 million date palms in the Kingdom of Saudi Arabia and ranks third on the globe in date production (FAOSTAT, 2017). Date palms are threatened by an invasive pest, the red palm weevil (RPW), Rhynchophorus ferrugineus. The number of date palms infested by RPW is high; an estimated number of 80,000 infested palm trees is reported in Saudi Arabia \cite{dawood28-mukhtar2011new}, and numbers are rising daily. The RPW is reported to be a category-1 pest to date palms in the Gulf area \cite{dawood11-el2009threat}.
The larvae of RPW are the causing agent in the infestation process, where they feed internally in the date palms trunks. This feeding behavior delays early detection. Failure in early detection of the infestation causes rotting of the internal parts of the palm trunk leading to its death \cite{dawood3-abraham1998integrated}. The RPW can complete many generations a year in the same host before its death \cite{dawood31-rajamanickam1995certain};\cite{dawood7-faghih1996biology}.
Early detection of RPW infestation is extremely hard because neither RPW larvae nor damage can be observed at this stage.  The following Infestation symptoms are visible at later stages: (1) presence of galleries in the trunk of the tree, (2) oozing out of brownish viscous material with a fermented odor, (3) collection of chewed fibers between leaf bases around the trunk, (4) hearing of the feeding larvae sounds when the ear is placed close to the palm trunk, (5) presence of empty RPW cocoons and dead adults underneath the infested date palm tree, and finally (6) the falling of date palms by breaking down of the palm trunk parts \cite{dawood19-hussain2013managing}.

So far, different methods have been tested to detect the RPW infestation in date palms in early stages. These include visual inspections and acoustic sensors \cite{dawood29-potamitis2009automatic}, sniffer dogs \cite{dawood27-nakash2000suggestion}, pheromone traps \cite{dawood12-faleiro2008rapid}, transcriptome analysis \cite{dawood15-giovino2015transcriptome}, Laser Induced Breakdown Spectroscopy (LIBS) \cite{dawood14-farooq2015application} and Near Infrared Spectroscopy (NIRS) technique \cite{dawood22-kemsley2008feasibility}, thermal camera, radar 2000, radar 900 and resistograph; but none could prove desired results. Scientists are still making efforts to discover some effective, efficient and environmentally safe method for RPW early detection \cite{dawood2-abdullah2009biological}.

Visual examination of a tree is one of the regular methods that farmers perform by looking for heavy brown liquid on the palm trunk or the existence of regular or semi-regular holes on the trunk \cite{tofailli2010}. Pheromone traps are also widely used to determine the presence of RPWs in certain areas, although this method does not specify the exact infected palms.  \cite{Faleiro2008}. For best results, the pheromone traps need to be examined regularly including a collection of the weevils, cleaning of the traps and replacement of the exhausted pheromone. The visual inspection and pheromone trapping, as well as the other traditional approaches, are the main techniques used currently for RPW detection and monitoring. However, many limitations are facing these techniques that include: financial difficulties, trained human resources, and farmer's conception that RPW adults are attracted to their healthy farms through the presence of pheromone traps. Moreover, the poorly maintained and, neglected old orchards and date palm plantations on study side and in parks are not under the surveillance programs which sometimes act as a source of a secondary infestation.

Technological techniques are emerging in response, such as acoustic sensors \cite{Potamitis2009}. Acoustic technology measuring the spectral and temporal patterns of sounds produced by feeding and moving larvae has the potential to enable early detection, mainly because the insect sounds often can be distinguished from agricultural or urban background noise.

Currently available acoustic systems have seen limited use because they require skilled operators. Near-Infrared Spectroscopy (NIRS) technique \cite{KEMSLEY2008223} has been extensively used for non-destructive analysis and monitoring of biological systems. In NIRS, the specific chemical composition of an object excites molecules to absorb light in the NIR region and vibrate at unique frequencies. Insect borers cause stress to the plants interfering with transpiration stream by ingesting plant stem tissues. Similarly, when RPW infest date palm, it starts eating internal tissues of the tree and induces stress that can be detected through the NIRS technique. A preliminary study was carried out in Saudi Arabia by measuring absorbance spectra for control, wounded and RPW infested fresh date palm leaves samples through spectrophotometry. Preliminary results are promising and provide evidence that the NIRS technique has the potential for RPW detection at early stages. Infrared cameras are in use to detect temperature increase in infested palms. Currently, available literature on this aspect suggests that baseline information on temperature profiles of RPW infested date palms are available for developing a real-time sensor. Two models of IR Thermal Cameras were tested in the field in different seasons (summer and winter) to assess their efficacy in identifying the RPW damaged palms. When the thermographs of healthy and damaged palms were analyzed in some cases, the differences in the color spectrum was clear and easy to mark the damage. However, it was not easy when the surface temperature and inside temperature were not much different.

The Laser-induced breakdown spectroscopy (LIBS) technique was applied on the soil surrounding the trunk of the RPW infested date palm \cite{Farooq2015}. Results of this study showed that the presence of different elements; such as Ca, Mg, Na, C, K, and OH and CN molecules; could be used as indicators of possible differentiating factors between infested and healthy samples. The study showed that the Mg and Ca atomic lines intensity in LIBS spectra increases rapidly with the growth of the population of the pest. These results indicate that the LIBS technique as a non-destructive method has the potential to be used as early detection for RPW infestation.

The X-Ray is a widely used technique for medical imaging, but, its use in agriculture for the detection of insect pest infestation is relatively less. Preliminary studies carried out under laboratory conditions revealed promising results. Results indicated images of the larval stages and the galleries created by RPW larvae inside the date palm trunk. Further studies are needed to improve the methodology for imaging the tree and devising system compatible with date palm tree imaging under field conditions.

Near-infrared detection experimentation will hasten the use of drone-based early detection if such sensors are made based on the results of these experiments. Also, experimentation for the creation of a portable Laser-induced breakdown spectroscopy-based technology would be a handy tool in the early detection of RPW on the ground. Furthermore, in this matter, high-frequency radar and X-ray technology experiments have some promises based on preliminary experiments.

Besides, proteomics methods have been widely used for detecting human infections and diseases. A few plant-related proteomic studies would encourage the utilization of these methods for detecting RPW infestation at early stages of date palms infestations. The production of diagnostic molecular markers to be used as early detection of RPW infestation tools would work if these protein molecules would show a modulated response. This response can be used if changes in these date palms are linked with the presence of infestation by RPW. Also, experiments with proteomics strategies carry a high potential for developing future kits for early detection.

Based on the above, it can be concluded that none of such technological techniques prove desired results. Scientists are still making efforts to discover some effective, efficient, and environmentally safe method for RPW early detection. The use of a framework that collects different data from multiple sensors would help in early detection success of RPW infestations.

If an infested palm is at discovered early, the efficacy of treatment is higher; which raises the significance of timely detection of RPW infestation. As long as the heart of the palm is not yet damaged and the trunk is still stable, the palm can be treated and usually recover \cite{Soroker2013}. The RPW treatment by preventive physical, biological, or chemical means is still under research \cite{GUTIERREZ2010}. However, severely infested trees, once detected, must be destroyed \cite{Potamitis2008} and completely removed as no efficient treatment process currently exists \cite{Hussein2009}. 

In \cite{s130201706}, the authors have proposed a bioacoustic sensor for the early detection of Red Palm Weevil. It consists of a sound probe, a microphone, a processor for local computation, solar panel for energy, and a wireless interface for sending the data to a control station. Extensive experiments showed that the sensor can detect with an accuracy of 90\%.

\section{ Smart Palm Architecture}
In what follows, we present the general system architecture of the smart palm system and we discuss in details its different components. 

\subsection{General system architecture}
The system is decomposed into different layers. 
\begin{figure*}[htb!]
	\centering
	\includegraphics[width=15cm]{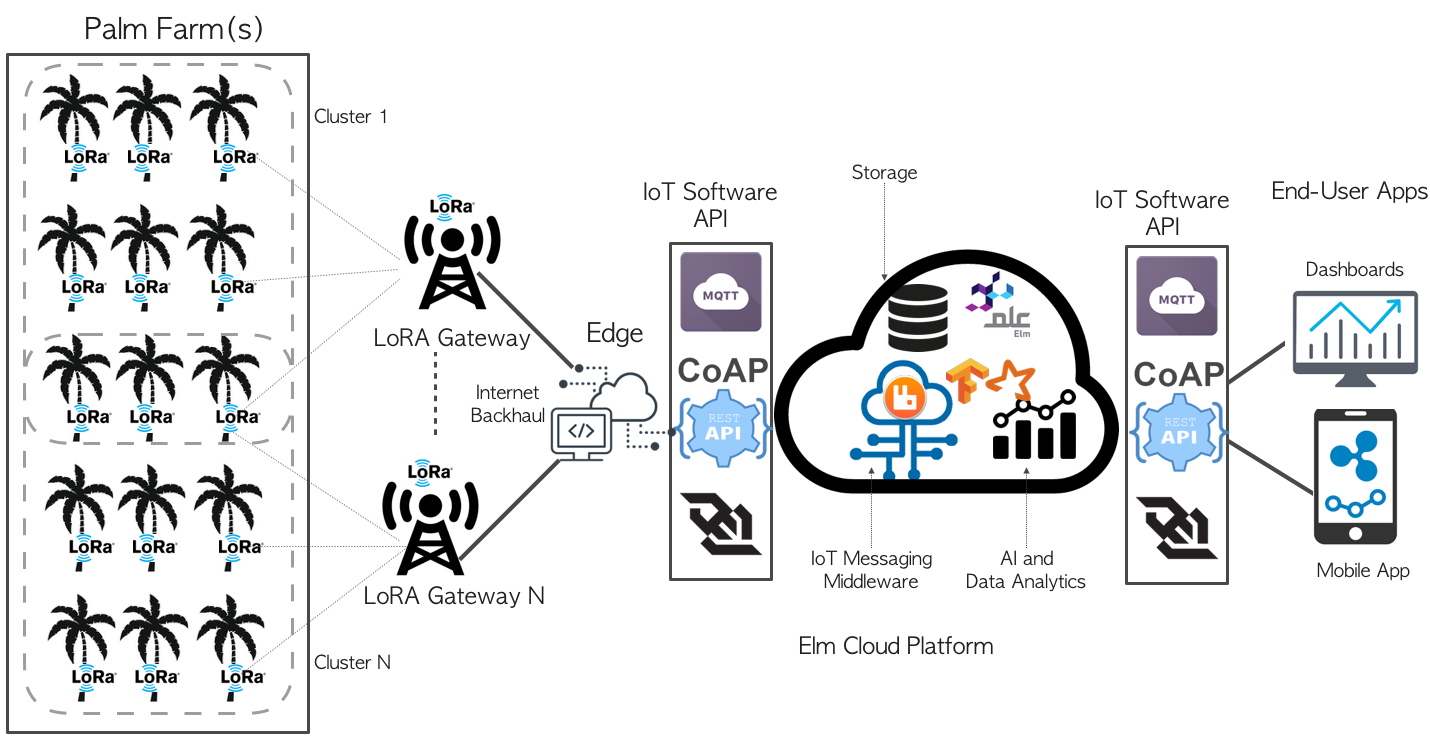} 
	\caption{Smart Palm Architecture}
\end{figure*}

\begin{itemize}
	\item The Palm Farm Layer: This layer is the data source layer that contains all the palms distributed in multiple farms. Every palm is equipped with a sensor that captures vital information of the palm, including temperature, humidity, PH level, in addition to weevil detection sensors, namely, a sound sensor (i.e. microphone) and a vibration sensor. This information is sent periodically from the sensor node to the cloud system through the gateway. The sensor nodes are equipped with a LoRA shield to allow very long range communication with the gateway. Several studies show that LoRA achieves distances up to 20 Km, but with low data rates. In the smart palm application, the data rate is very small as data contain small number of bits and will be sent after long periods of time. 
	As a farm can be very large, it is may be decomposed into several clusters of palms, where every cluster connects to one particular LoRA gateway. In fact, a LoRA gateway can handle a limited number of connections, thus, it is necessary to have multiple cluster for farms of very large sizes. The configuration of the number of clusters per farm is a design choice.
	\item The LoRA Gateway Layer: this layer is the collection point of the data coming from different palms. Given that the LoRA communication range is very large, the gateway will collect data from a large number of nodes, and then forwards them to the next Edge layer. The gateway is just a simple forwarder of messages. It does not perform any kind of processing on the data. It receives the palm vital data from its LoRA interface and forwards it to the edge through its IP Interface. Usually, both Ethernet and WiFi can be used to transmit data to the IP network.
	\item The Edge Layer: Data coming from palms should be transmitted to the cloud. However, as the number of palms increases, which can reach millions of palms, sending all data collected from palms directly to the cloud leads to higher storage requirements at the cloud, in addition to a higher computation complexity. This approach would lead to high management costs at the cloud in addition to possible degradation of the quality of service due to increased latency. To overcome this problem, we use an edge layer, which is an intermediate layer between the farm system and the cloud, and the edge will be responsible for a subset of data, will do local processing and send only a digest to the cloud through aggregation, thus reducing the load on the cloud and improving the QoS through reduced latency as it is closer to the devices. Edge are typically distributed in nature, and they are located in a region close to where the data is originated from. For example, assuming that we have 1 million of palms across the country and each palm sends one message every one minute.  This will result into 3.6 billion messages per day sent to the cloud. This of course will be require extensive storage capability and computation as well. Instead of sending data to the cloud, if we assume that we have 100 edges, then these edge will receive and locally store and process the information, and will send a summary of these data (i.e. max, mean, min, ...) periodically (for example, every 1 hour) to the cloud. This solution is more scalable and more energy efficient from the cloud perspective. An edge can be a local server or machine specific to a region or a city. For small systems, the edge layer can be ignored and directly transfer data to the cloud. 
	\item The Cloud Layer: The cloud layer is the main central system that ensure storage and processing of collected data and also the communication between the different entities of the system. The cloud layer contains a messaging middleware system (Kafka, RabbitMQ) that allows to exchange messages between data sources and user applications. It can be access through different software API and communication protocols, namely MQTT, WebSockets, REST Web Service interfaces, and COAP. The cloud layer also contains big data analytics tool to perform computation and analysis of the collect data and implement advanced machine learning algorithms for the detection of the weevil infestation. It allows also to perform statistics on the data and provide useful insights to the farmers. 
	\item The End-User App Layer: The end-user applications layer contains the mobile and web applications that allow user to monitor in real the status of the their farms remotely. User friendly dashboard provides the users with a comprehensive view about the state of their farms, namely, the number and location of infected palms, insight on the propagation of the infestation over time, in addition to the vial data for the palms. The user can select information for a particular palm either in real time, or the historical. He can also display information about the whole farm and particular clusters.
	
\end{itemize}

\subsection{Software Architecture}
\begin{figure}[htb!]
	\centering
	\includegraphics[width=9cm]{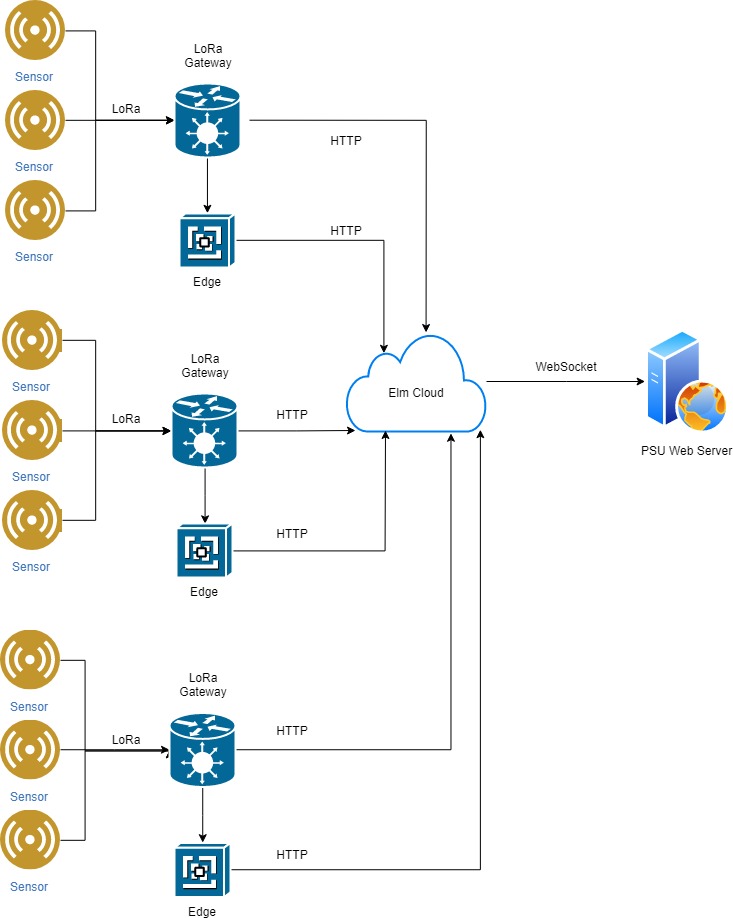} 
	\caption{Flow diagram of the software system}
	\label{flow}
\end{figure}

In this section, we present the software architecture of the system. The UML flow diagram presented in Figure \ref{flow} shows the interaction between the components of the system.

The process starts when the nodes receive the data from their sensors. The nodes forward the data to a gateway through their LoRa protocol interface. Then, the gateway sends the data to two destinations. The first destination is the edge where local computation is performed to get intermediate results, as explained in the general architecture section. These results are then sent to the cloud. The second destination is the cloud itself to keep the data stored for later usage. The edge receives the data from the gateway through a WebSocket interface, while the cloud will receive them through HTTP or CoAP. However; since the gateways are not constrained devices, the data is sent through an HTTP request. Finally, the cloud sends the data received from the sensors through a WebSocket to the Web Server, which in turn sends them to the client side after doing some filtering based on the client requirements.

\begin{figure}[htb!]
	\centering
	\includegraphics[width=9cm]{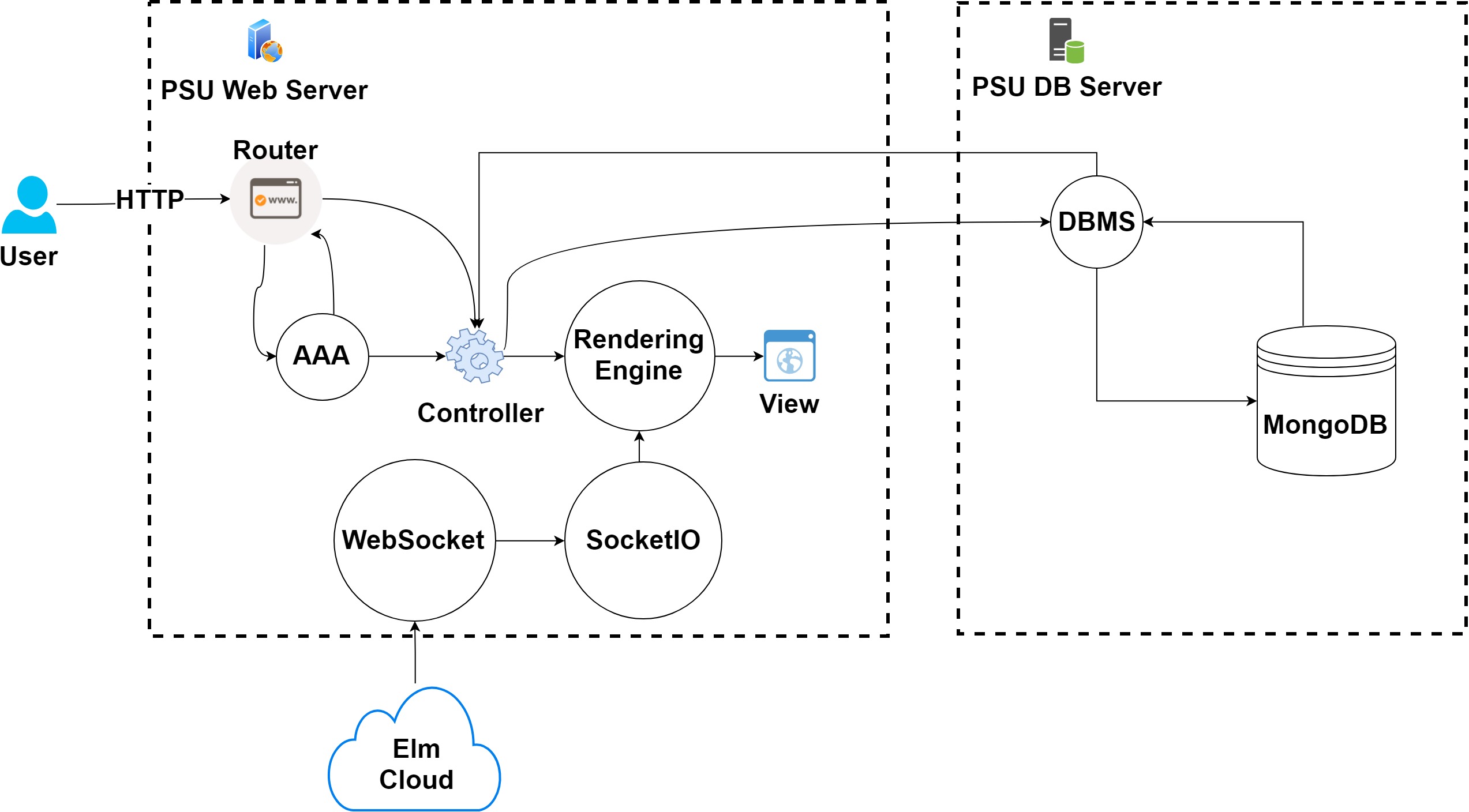} 
	\caption{Flow diagram of the Web Server and the client-Website}
	\label{flow2}
\end{figure}

Figure \ref{flow2} represents the UML flow diagram that shows the interaction process between the Web Server and the client website. 

The data from the nodes is received from the cloud through a WebSocket interface, then in the WebServer, we are using Socket.IO to send the data to the client application. Socket.IO is a library to transform the data through different protocols based on the situation at hand, but mainly through a WebSocket. 
When a user requests the server to open a page, his request is intercepted by the \textit{Router component}. The Router redirect the users to the requested page. However; the request is first checked by the Authentication, Authorization, and Accounting component (AAA component). If the page contains sensitive or user-based data and the user is valid and authorized to access this page, the requested data is sent to the controller which processes the request and returns the requested data. The data is rendered to the user browser through a rendering engine that handles the views that the user should see. In addition, when the user performs an action that its results or the action itself needs to be logged, these new data is stored in a database for later auditing and analysis.

\begin{figure}[htb!]
	\centering
	\includegraphics[width=9cm]{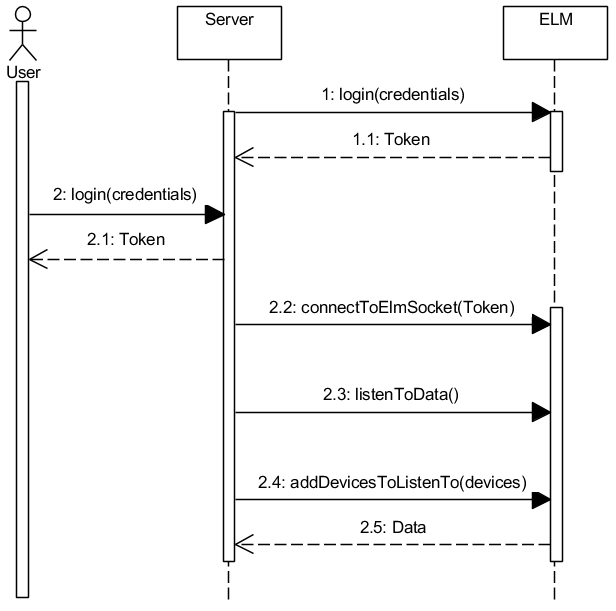} 
	\caption{Sequence diagram of the Web Server, the cloud and the user}
	\label{sequence-user-server-cloud}
\end{figure}

Figure \ref{sequence-user-server-cloud} presents the UML sequence diagram for the interaction between the user, the server and the cloud. First, the Web Server must be authenticated with the cloud, which in our case is the ELM cloud platform. 
Then, the server connects to the cloud through a WebSocket interface and listens to the data, and it sends to the cloud the list of devices from which it wants to receive their sensor readings in real-time. In response, the cloud starts sending the readings of these specified devices continuously in real time to the server. Finally, the user can view these data by providing his credentials to be able to open the dashboard.

\begin{figure}[htb!]
	\centering
	\includegraphics[width=9cm]{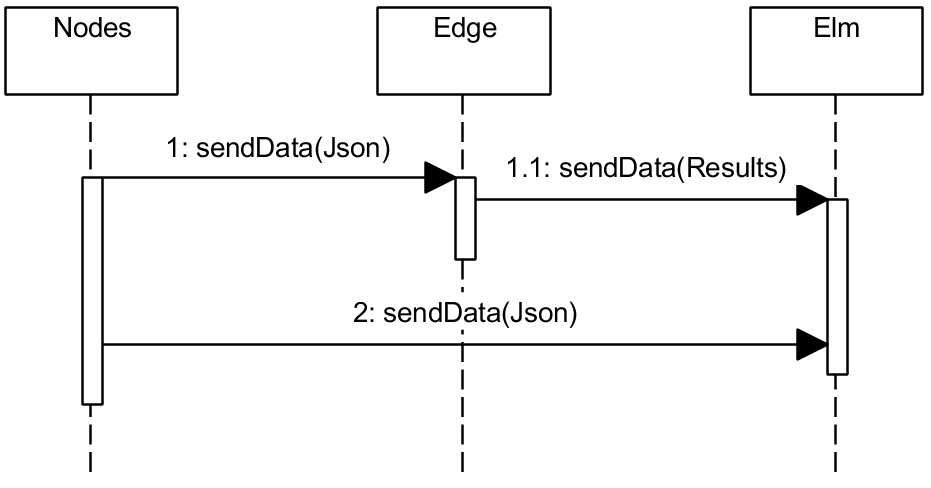} 
	\caption{Sequence diagram of the nodes, edges and the cloud}
	\label{sequence-node-edge-cloud}
\end{figure}

Figure \ref{sequence-node-edge-cloud} represents the sequence diagram that shows the order in which the data is traveling between the nodes, edges, and the cloud. The nodes send all the data without an exception to both the edge devices and the cloud. However; the edge machines perform local analysis on these data and finally send the results to the cloud which stores them to be used later for further analysis. The reason for why we are sending all the data from the nodes not only to the edges but also to the cloud although the edge is the entity responsible for performing local analysis and providing the results is that the cloud will store all of these data for historical analysis and trying to discover more accurate methods to discover the Red palm weevil, while the edges only apply the methods chosen based on the cloud analysis to reduce the load on the cloud.

\subsection{Web Client Application}
In this section, we present the Web client interface that allows users to monitor palm farms and receive vital information of the palms in addition to alarms on detected red weevil palm infestation. 

\begin{figure}[htb!]
	\centering
	\includegraphics[width=7cm]{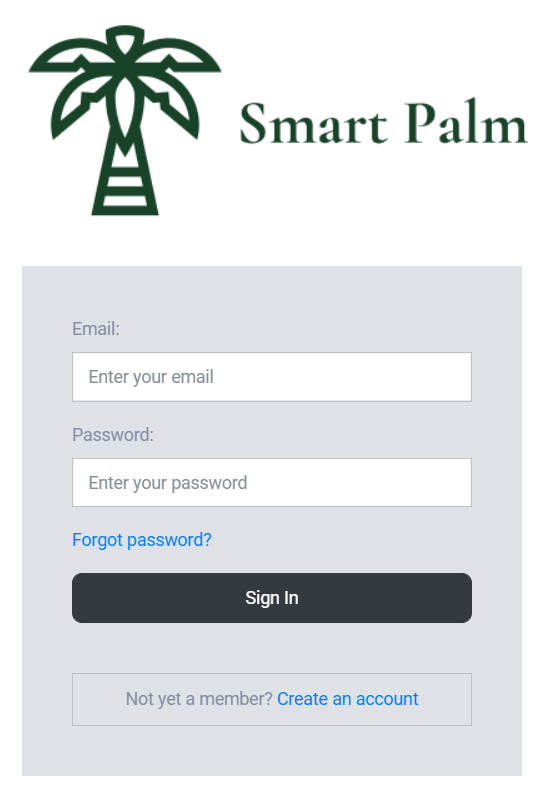}
	\caption{Sign-In Page}
	\label{signin}
\end{figure}

The client application includes several features that makes the process of monitoring the palms status effective and user friendly. When the user opens the website for the first time, a sign-in page is shown up as shown in Figure \ref{signin}.

\begin{figure}[htb!]
	\centering
	\includegraphics[width=9cm]{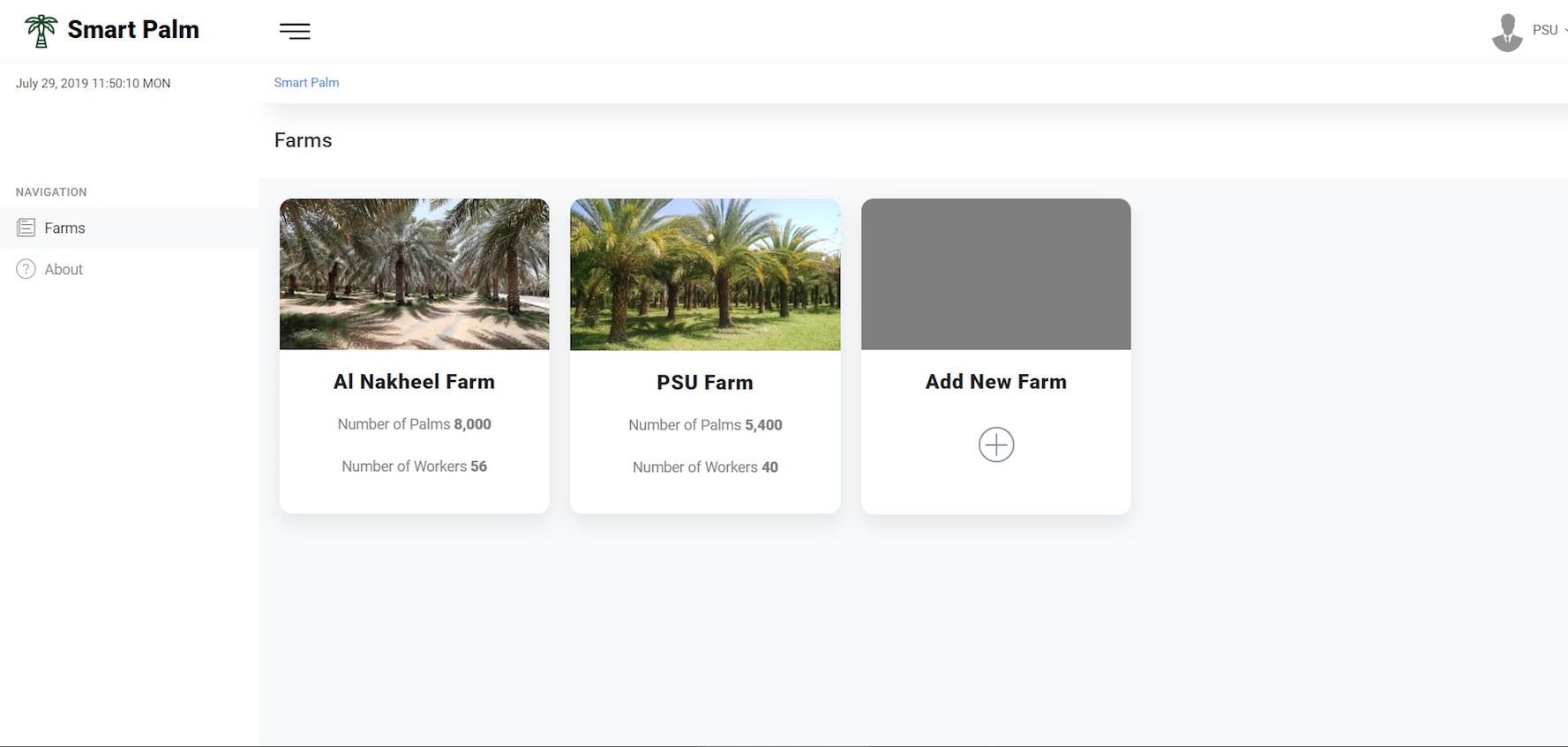}
	\caption{User's Farms Interface}
	\label{farms}
\end{figure}

After the user provides his credentials and clicks the Sign-In button, a new page opens and shows the farms that the user has access to, then he can select the one he wants to view as shown in Figure \ref{farms}.

\begin{figure}[htb!]
	\centering
	\label{fs4}
	\includegraphics[width=9cm]{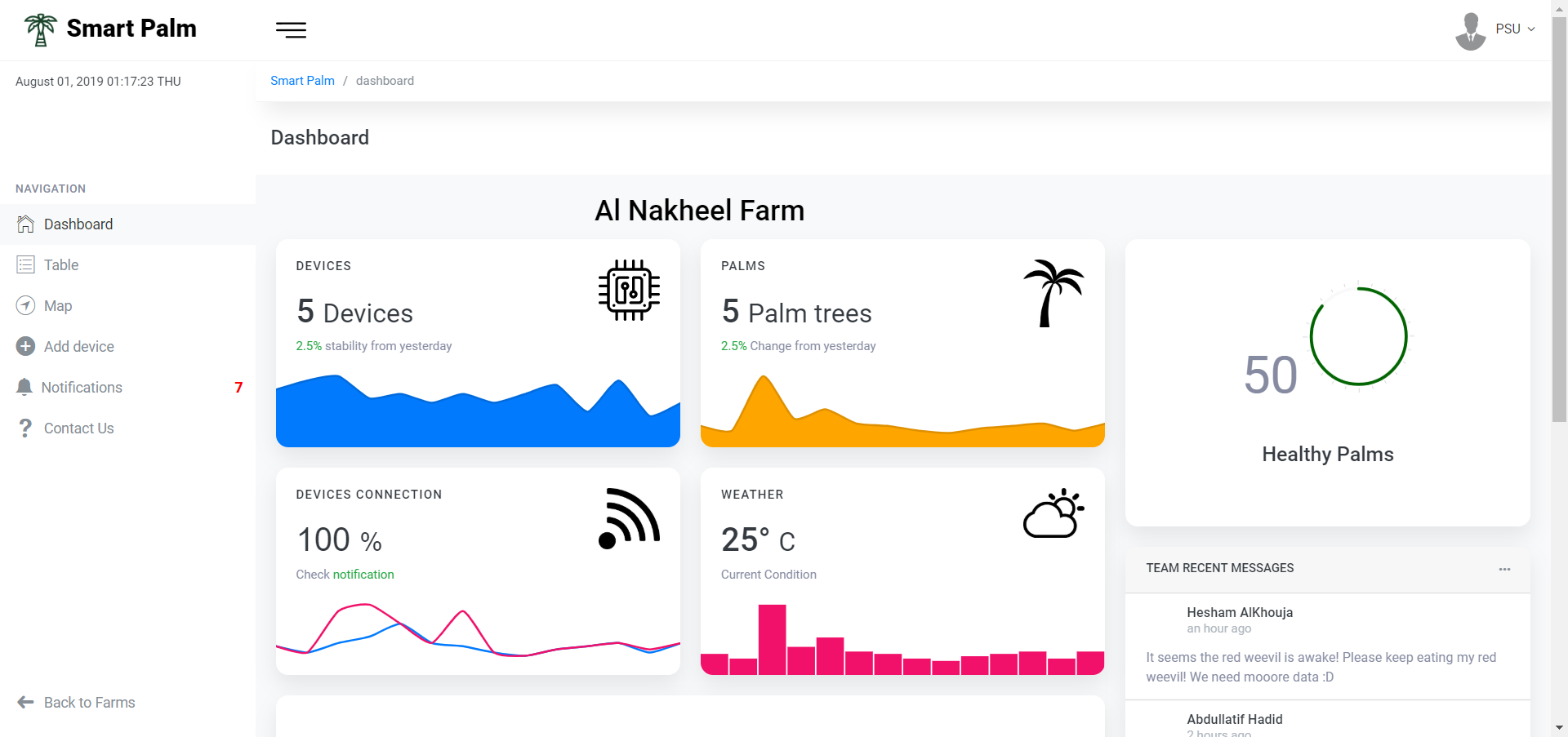}
	\caption{Dashboard main page}
	\label{dashboard}
\end{figure}

\begin{figure}[htb!]
	\centering
	\label{fs4}
	\includegraphics[width=8cm]{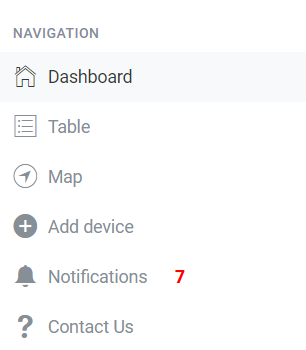}
	\caption{Side navigation menu}
	\label{sidepane}
\end{figure}

After selecting a farm, a dashboard page opens, as shown in Figure \ref{dashboard}. The page provides details and analytics about different aspects related to the selected farm, including the number of palms, the percentage of healthy palms, the weather and other useful statistics. The page also has a side navigation menu at the most left part of the view, shown in Figure \ref{sidepane}. This menu allows the user to select the page based on the information/action he wants to view/perform.

\begin{figure}[htb!]
	\centering
	\includegraphics[width=9cm]{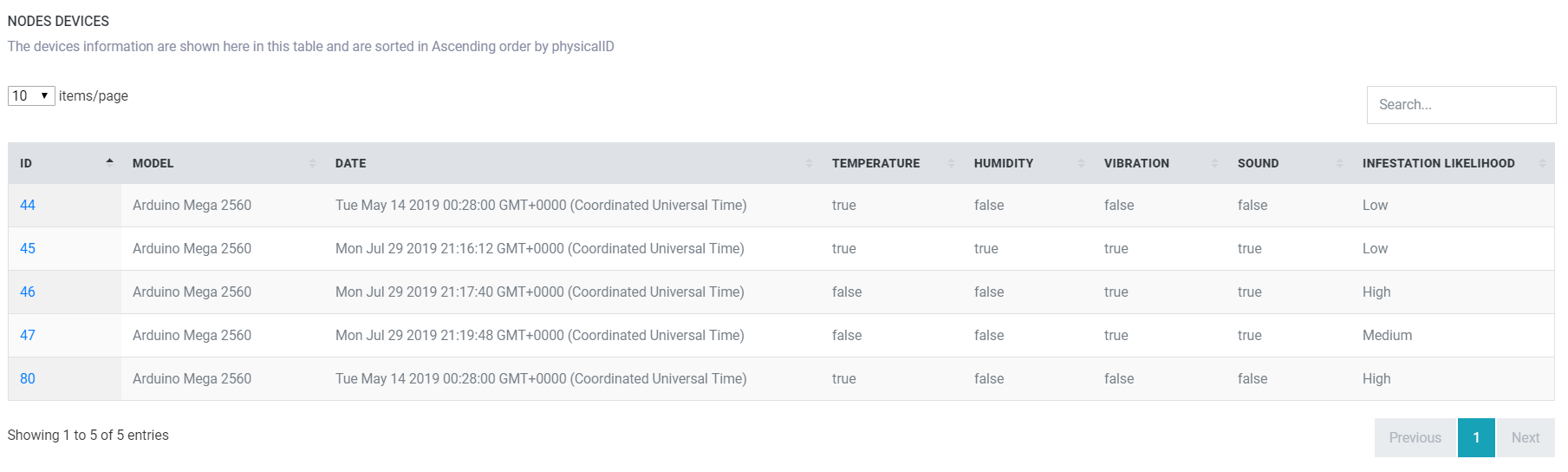}
	\caption{devices - table page}
	\label{table}
\end{figure}

As shown in Figure \ref{table}, the user can navigate to the table page, which displays the palm's general information as a data grid.

\begin{figure}[htb!]
	\centering
	\includegraphics[width=9cm]{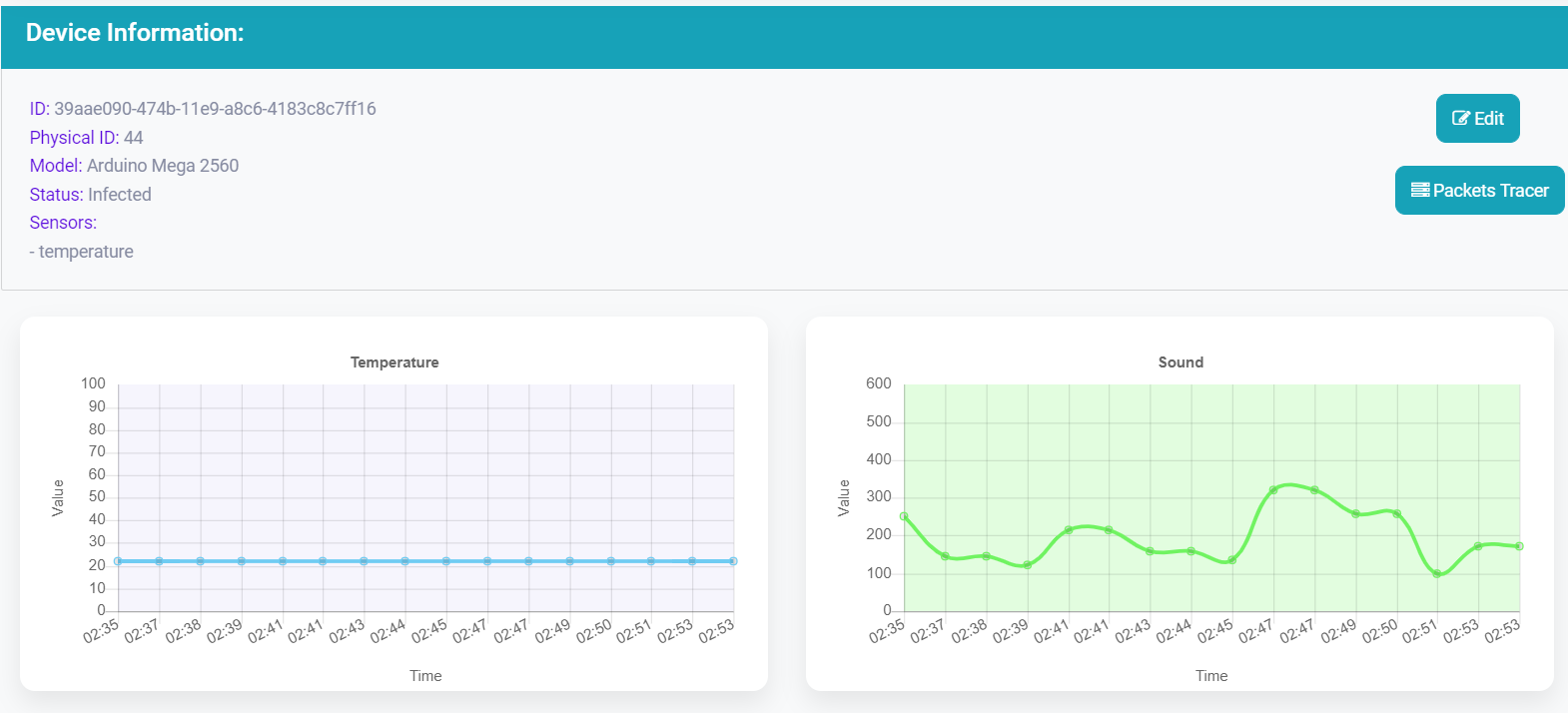}
	\caption{Real-Time Data Representation - 1}
	\label{device-details-1}
\end{figure}

\begin{figure}[htb!]
	\centering
	\label{fs4}
	\includegraphics[width=9cm]{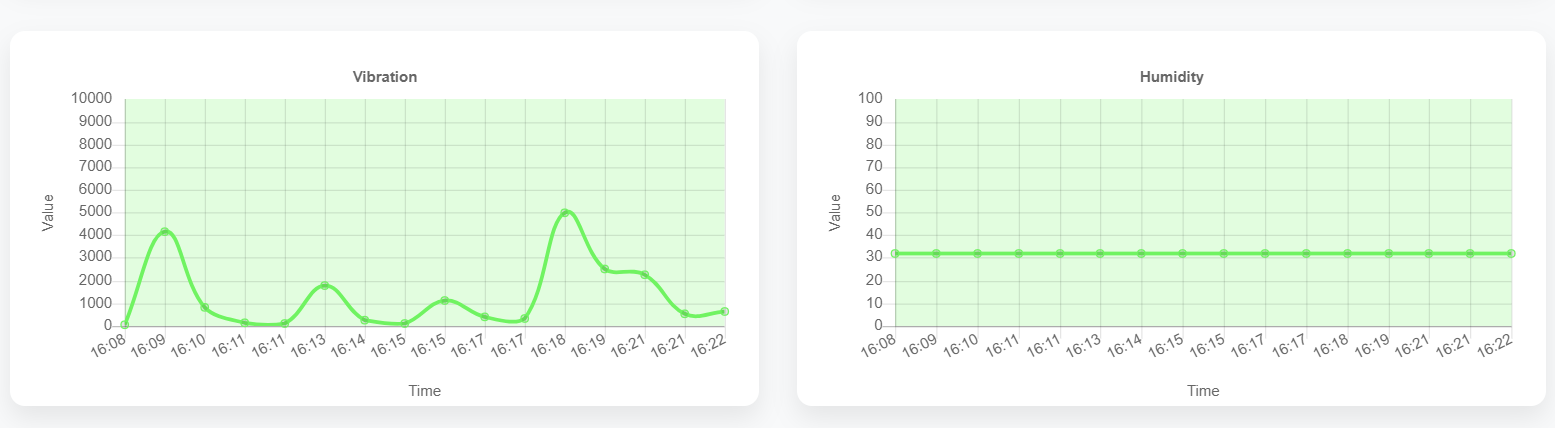}
	\caption{real time data representation - 2}
	\label{device-details-2}
\end{figure}

Each row represents a palm as well as the device attached to it. The table clearly shows the infestation likelihood in three different levels, high, medium, and low. and this value is updated in real-time. The dashboard also shows the sensors that are attached to each device. Moreover, it has a search box that helps in searching the table based on specified parameters (e.g., the ID of the device). When an ID number of a device on the table is selected, a new page opens and shows more detailed information about the palm/device, as shown in Figures \ref{device-details-1}-\ref{device-details-2}. This page also gives the user the ability to edit the selected device. Besides, it shows different graphs based on the sensors attached.

\begin{figure}[htb!]
	\centering
	\includegraphics[width=9cm]{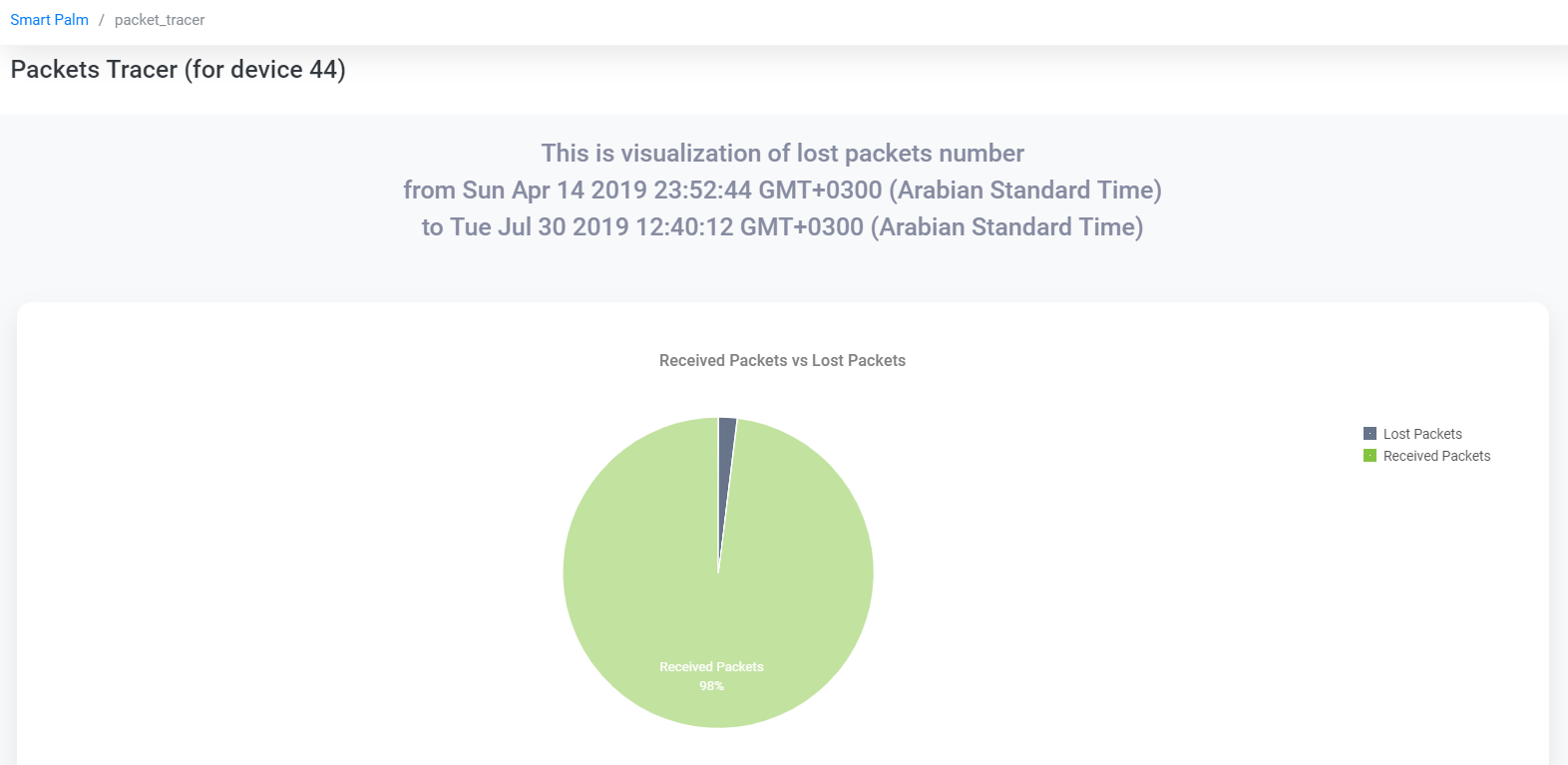}
	\caption{packet tracer - 1}
	\label{packet-tracer1}
\end{figure}

\begin{figure}[htb!]
	\centering
	\includegraphics[width=9cm]{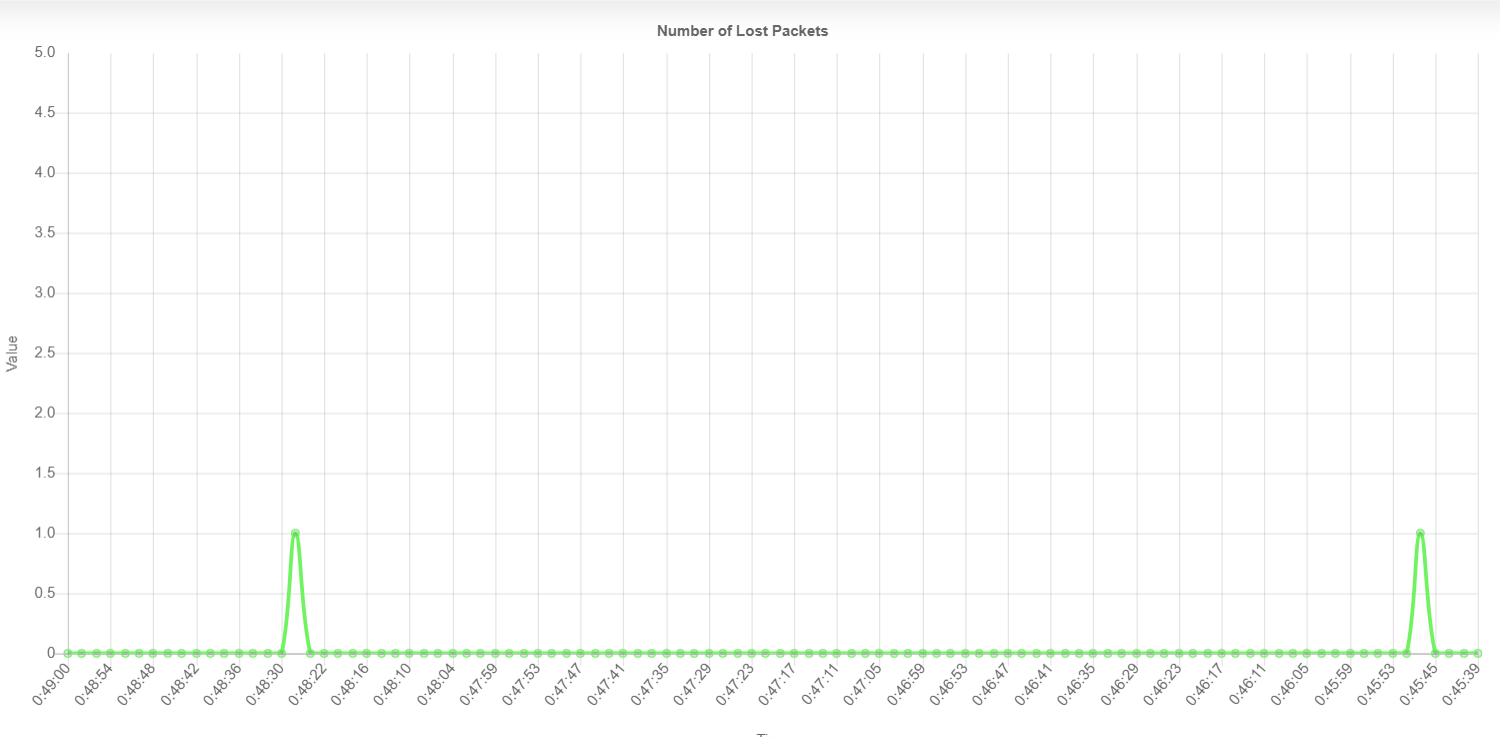}
	\caption{packet tracer - 2}
	\label{packet-tracer2}
\end{figure}

These graphs show the flow of data out of each sensor in real-time. This page also has a packet tracer button that navigates to a page that shows the percentage of the packets received compared to the lost packets in a specific range of time [Figure  \ref{packet-tracer1}-\ref{packet-tracer2}]. It helps in identifying if there is an issue in the connectivity to the device.

\begin{figure}[htb!]
	\centering
	\includegraphics[width=9cm]{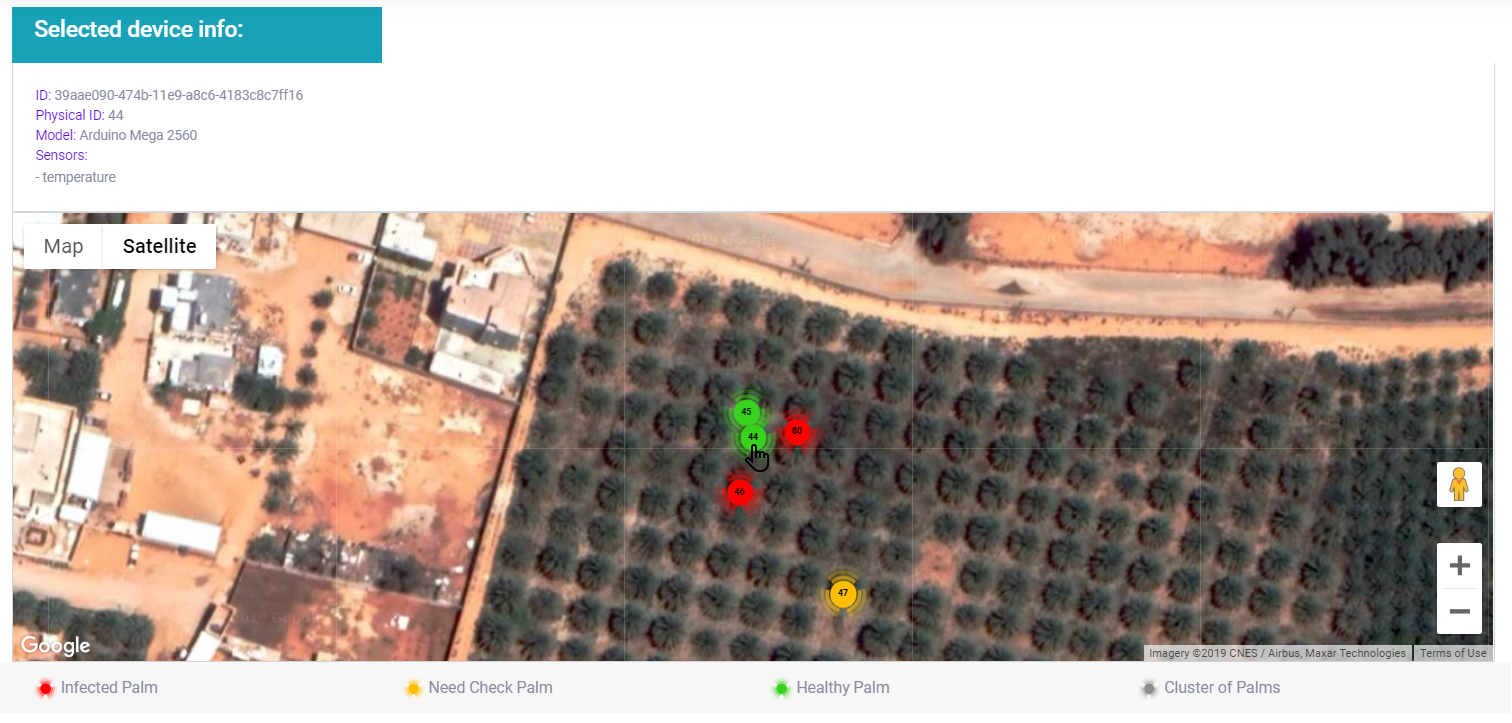}
	\caption{Palms map. The green icons represent healthy palms, the red icons represent infected palms, and the yellow icons represents a palm the require attention.}
	\label{map1}
\end{figure}

\begin{figure}[htb!]
	\centering
	\includegraphics[width=9cm]{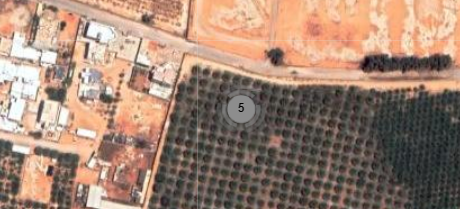}
	\caption{Palms map - The figure shows a cluster view of five palms}
	\label{map2}
\end{figure}

As for the side navigation menu, the user might want to open the map page as an alternative way other than the table view. The map view is user-friendly. It shows each palm on the map accurately based on the latitude and longitude of each device [Figure \ref{map1}-\ref{map2}].

\begin{figure}[htb!]
	\centering
	\includegraphics[width=9cm]{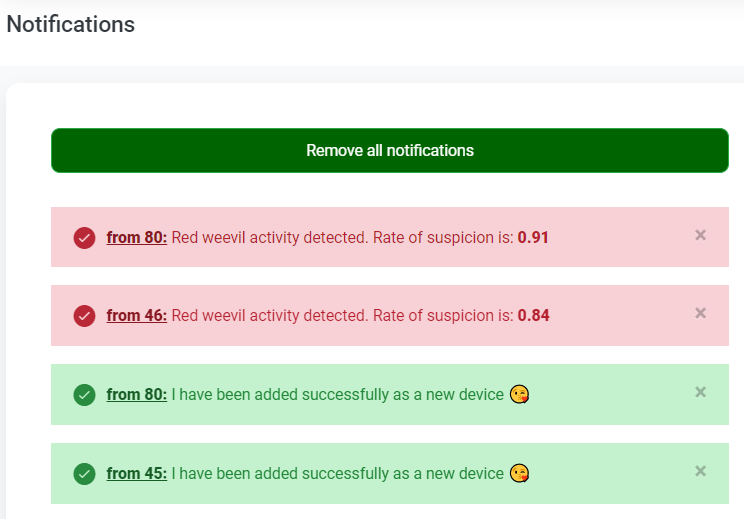}
	\caption{Notification page}
	\label{notifications}
\end{figure}

The map view distinguishes each palm state by a different color. The green color refers to a healthy palm, the red color to an infected palm, and yellow conveys the suspicious of infection. The grey color appears when the map is zoomed out, and the palms are clustered in one single icon. This makes the view more compact and provides the number of palms in that area. When the user hovers a palm icon on the map, the palm/device info is shown in the selected device information box on the same page. When the user clicks on the icon, it navigates to the corresponding page mentioned above when the user clicks on an ID in the table page. Coming back again to the side navigation menu, the user can select the add device option to add a device manually and specify the different sensors attached.
Nonetheless, this is not recommended. It is better to let the LoRa gateways detect the devices in their region and add them automatically. The next option in the side navigation menu is the Notification page [Figure 17]. The side navigation menu shows the number of notifications besides the Notification option to alert the user of the existence of new notifications [\ref{notifications}].
The Notification page lets the user be aware of changes in the palms’ infection status as well as the detection and addition of new devices.

\section{Data Collection and Analysis}

\subsection{Experimental study}
In this section, we present the experimental study that consists in using different sensors to detect the weevil insect in a palm tree. Then, we present the data analysis, discuss the findings and present the final recommendations. 

First, we present the experimental settings of the different sensors used for weevil detection, and then we explain the data collection process. We also present the different experimental scenarios, the type of sensors used and the different parameters of the collection process. Second, we present different types of performance metrics used for weevil detection. Then, we analyze and compare the results using the proposed metrics. Finally, we discuss the findings and provide some recommendations based on the results. 

\subsubsection{Experimental settings}

\paragraph{Sensor selection}

The selection of the appropriate sensor for the weevil detection was a tedious and challenging process. In fact, the vibration and sounds generated by the weevil are found to be very small and minimal, which makes their detection through the palm tree a challenge. 

We have tested around 7 types of sensors ranging from vibration sensors, accelerometer sensors and sound sensors. Among them, the accelerometer sensors were found to be the most effective. 

\begin{figure}[htb!]
	\centering
	\includegraphics[width=9cm]{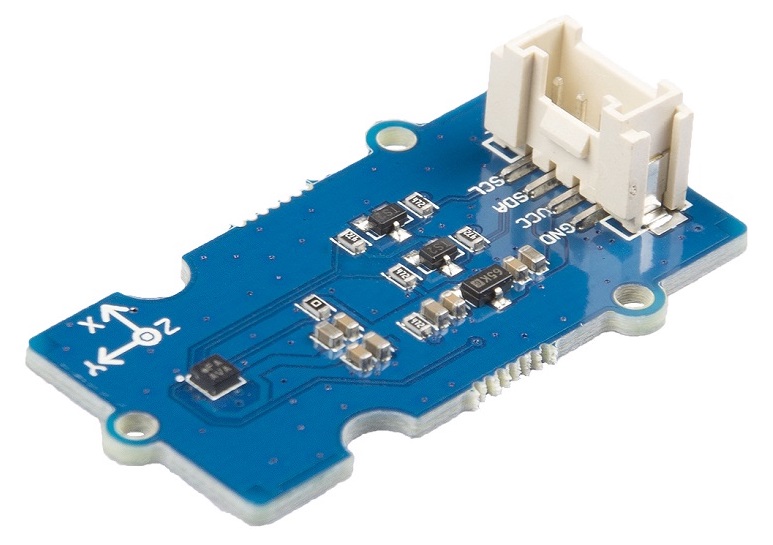} 
	\caption{Grove 3 Axis Digital Accelerometer ±16g Ultra-low Power BMA400 Sensor}
	\label{grovesfig}
\end{figure}

After investigation of these sensors, the experimental data collection was done using the sensor Grove 3 Axis Digital Accelerometer ±16g Ultra-low Power BMA400 as shown in Figure \ref{grovesfig}. 

The specification of the Accelerometer sensor is shown in Table \ref{table1}. Additional information about this sensor can be found at \cite{BMA400}.

\begin{table}[]
	\caption{Specification of the Grove 3 Axis Digital Accelerometer Sensor}
	\label{table1}
	\begin{tabular}{|l|l|}
		\hline
		Item                  & Value                                                                                                    \\ \hline
		Operating Voltage     & 3.3V/5V                                                                                                  \\ \hline
		Power Consumption     & \begin{tabular}[c]{@{}l@{}}18uA @5V \\ 14uA @3.3V\end{tabular}                                           \\ \hline
		Operating Temperature & -40℃ $\sim$+85℃                                                                                          \\ \hline
		Acceleration Range    & ±2g, ±4g, ±8g, ±16g                                                                                      \\ \hline
		Sensitivity           & \begin{tabular}[c]{@{}l@{}}1024LSB/g @±2g \\ 512LSB/g @±4g\\ 256LSB/g @±8g\\ 128LSB/g @±16g\end{tabular} \\ \hline
		Interface             & I2C                                                                                                      \\ \hline
		Size                  & L: 40mm W: 20mm H: 10mm                                                                                  \\ \hline
		Weight                & 3.2g                                                                                                     \\ \hline
		Package size          & L: 140mm W: 90mm H: 10mm                                                                                 \\ \hline
		Gross Weight          & 10g                                                                                                      \\ \hline
	\end{tabular}
	
\end{table}

It has to be noted that this BMA400 accelerometer sensor was selected among several other sensors, because it was found to be more sensitive than other tested sensors models, such as the Grove Vibration Sensor \cite{Piezo}, which was not sensitive enough to the vibrations made the weevil insect, and also the microphone sensors. In fact, the vibrations and sounds signals resulting from the infestation are very small and also do not propagate well through the thick trunk of the palm tree. However, the Grove 3 Axis Digital Accelerometer was found to be sensitive to the vibrations of the weevil insect activities much better than all other types of sensors that we have tested. 

\begin{figure}[htb!]
	\centering
	\includegraphics[width=9cm]{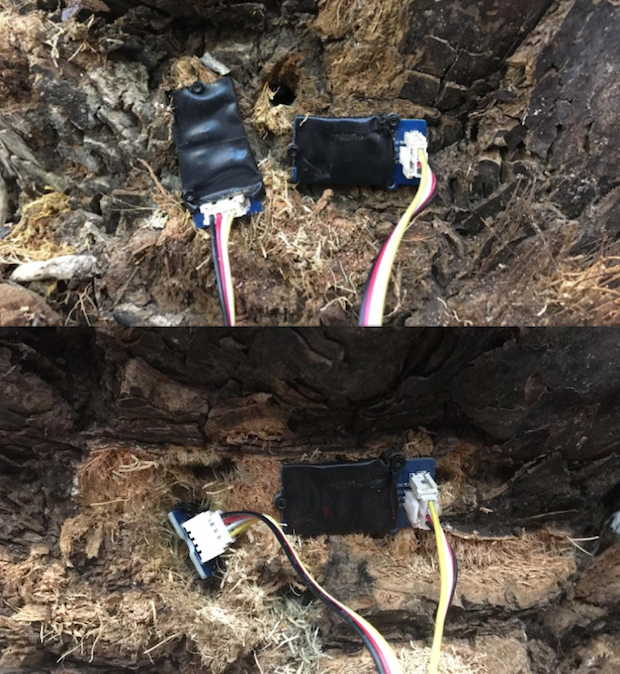} 
	\caption{Sensor Placement on the Palm Tree. The bottom figure corresponds to first scenario with one sensor inside and one sensor outside. The top figure corresponds to the second scenario with two sensors outside}
	\label{fig:sensor_in_palm}
\end{figure}

\paragraph{Data Collection Process and Scenarios}
The collection process was done at two different locations/heights that are close to the palm trunk base (around 1 meter above the ground). In fact, based on the feedback from experts, the red weevil infestation usually happens at the bottom of the tree.
We have used two sensors with different orientations at each location, as shown in Figure \ref{fig:sensor_in_palm}. 

In the first scenario, the sensor node is located at the height of 35cm and has two Accelerometer sensors. The first Accelerometer sensor was inserted \textit{inside} the palm on the depth of 4cm and second was put on top of the trunk after removing the old stems. The reason behind trying two different placements is to assess the sensor response when it is located inside the palm and when it is located on top of the trunk. 

\begin{figure}[htb!]
	\centering
	\includegraphics[width=9cm]{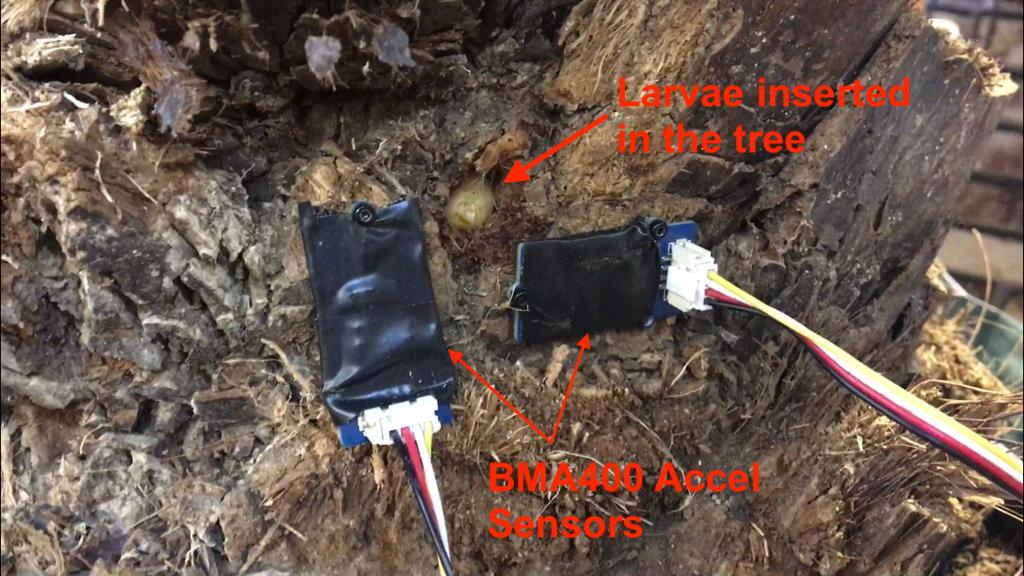} 
	\caption{Position of the sensors to detect Red Palm Weevil (Outside Sensors)}
	\label{fig:sensor_in_palm_2}
\end{figure}

In the second scenario, the sensor node was located at 20cm height. In this scenario, we put both Accelerometer sensors outside the palm tree, because we needed to collect more sensor data from outside the palm to check the consistency of collected data with the previous scenario. Figure \ref{fig:sensor_in_palm_2} illustrates scenario 2 and shows a larva just inserted inside the tree. 

\begin{figure}[htb!]
	\centering
	\includegraphics[width=9cm]{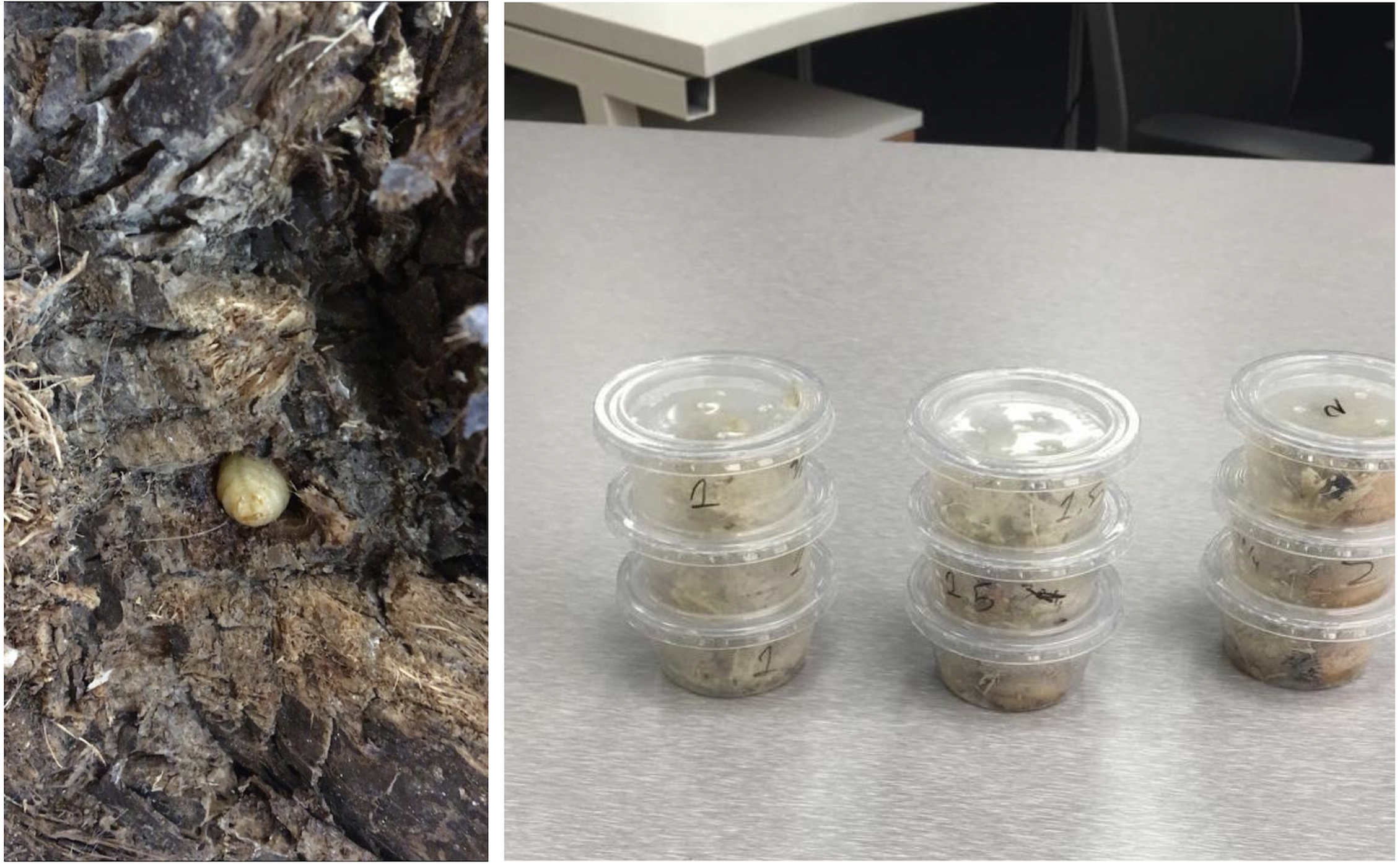} 
	\caption{Larvae used in experiments}
	\label{fig:LarvaeTree}
\end{figure}

The selected red palm weevil larvae were around \textcolor{blue} {1.5 to 2 months old}, because the larvae are most active at this stage before becoming a pupa. We inserted four larvae at the first location and \textcolor{blue}{one} at the second location. The data was collected and logged at both locations for six days before and after the insertion of the larvae to analyze the differences. \textcolor{blue}{In order to collect the baseline data, the data gathering process started before adding the Red Palm Larvae for 1 hour 24 minutes using a sensor outside the palm, and for two days using a sensor inside the palm. After inserting four larvae (three of 2 months old and one of 1.5-month old), the data was recorded inside and outside the palm for three days (2 days, 23 hours and 11 minutes). For phase two and after inserting a 1.5 months old larva, the data was recorded with two sensors outside the palm tree.  After collecting and logging the data, it was cleaned by removing the outliers, which are the values above 17g and the values under 6g. }\textcolor{blue}{The data was collected with a delay of 10ms (100Hz frequency),} and it included the acceleration on the X, Y and Z axes, their magnitudes, the date, the time and the number of the packets sent. Figure \ref{fig:LarvaeTree} shows a sample of larvae used in the experiments. 

The experimental settings used for data collection are summarized in Table \ref{table2}.

\begin{table}[]
	\caption{Experimental Settings for Data Collection}
	\begin{tabular}{|l|c|l|l|c|}
		\hline

		Location & \multicolumn{1}{l|}{Height (cm)} & Sensor (s) & Positions       & \multicolumn{1}{l|}{\begin{tabular}[c]{@{}l@{}}Number of\\  Larvae\end{tabular}} \\ \hline
		1        & 35                               & BMA400     & Inside, Outside & 4                                                                                \\ \hline
		2        & 20                               & BMA400     & Outside         & 2                                                                                \\ \hline
	\end{tabular}
\label{table2}
\end{table}

\subsubsection{Performance metrics}
The objective of this experimental study is to investigate the possibility of extracting signatures of the infestation of the red palm weevil insect from the data collected in the previously described experimental scenarios. For this purpose, we have analyzed the collected data using different techniques, namely signal processing methods, and statistical methods. In what follows, we present the metrics of interest. 
\begin{itemize}
	
	\item \textcolor{blue}{ \textbf{Signal processing techniques}: We used two approaches. The Fast Fourier Transform (FFT), which converts a time series from the time domain to the frequency domain. The objective of this technique is to investigate if the infestation of the palm would lead to different frequencies than in the case of a non-infested palm. Data were sampled at 100Hz. The second approach is the estimation of Power Spectral Density (PSD) using Welch’s method, which is also sampled at 100Hz and 2048-length for each segment. Hanning window was applied on both FFT and PSD to reduce leakage. To evaluate both FFT and PSD plots, the average value of the magnitude peaks is calculated based on the normalized threshold equal to 0.6.}
	
	
	\item \textbf{Statistical techniques:} For the statistical techniques, we have considered the box-plot representation, which summarizes six statistical values, namely the maximum and minimum of all the values, the median value, the standard deviation, and the 25\% and 75\% percentiles. We have also considered the cumulative distribution functions and compared them against each other for different scenarios. \textcolor{blue}{ Besides, we have studied the Histogram representation, which provides implicit distribution of the datasets.}
	
	
\end{itemize}

\subsection{Results analysis}
In this section, we analyze the collected data using signal processing techniques and statistical techniques. We consider the two cases where the sensors are deployed outside of the tree, and the case where the sensors are deployed inside the tree.\textcolor{blue}{ The data values were spited to hours approximately (instead of days) since RPWs cause minor changes on accelerometer readings, which make their signatures harder to observe in very large datasets. Data were collected for the first six hours. The plots were based on the combined acceleration through the X, Y, and Z axes all together (magnitudes). We considered the combined acceleration because all axes contribute to the detection process and it is independent of the placement of the sensor.}  

\subsubsection{Signal processing techniques}

\begin{figure*}[ht]
	\centering
	\includegraphics[width=15cm]{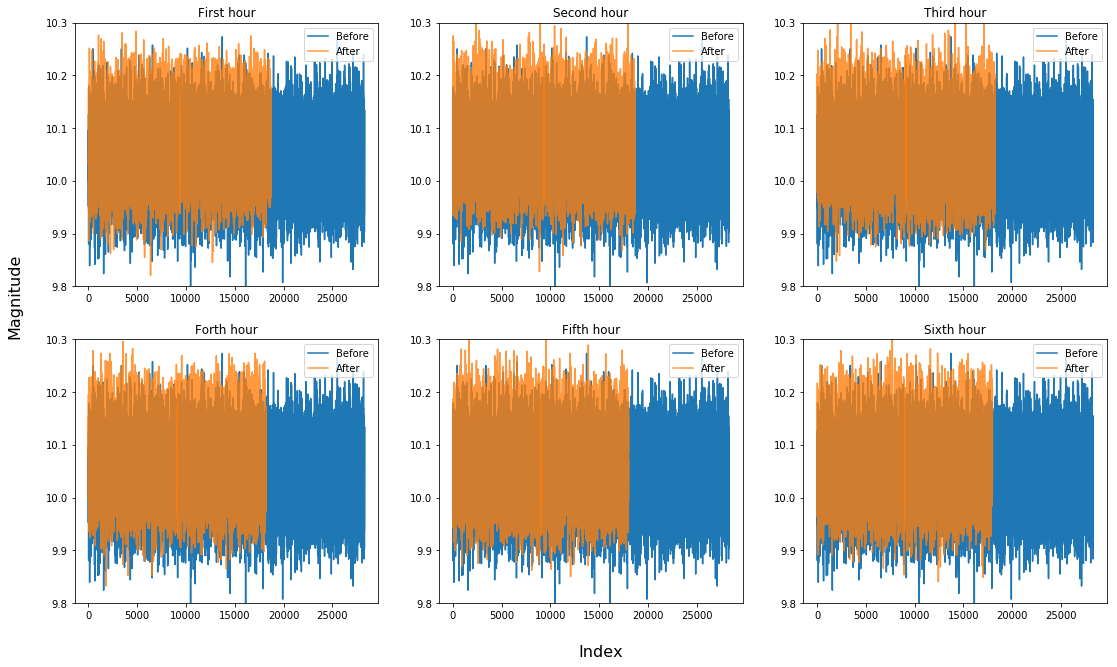} 
	\caption{Time domain plot for data collected from outside sensor before and after inserting Red Palm Weevils through hours}
	\label{fig:r_outside}
\end{figure*}

\begin{figure*}[ht]
	\centering
	\includegraphics[width=15cm]{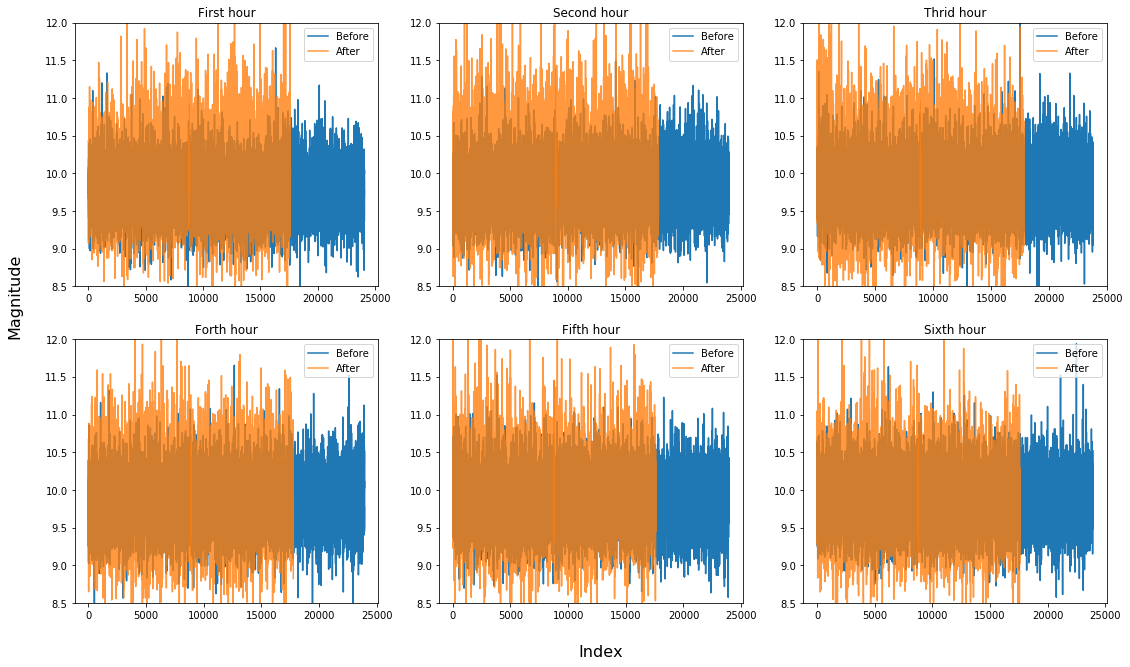} 
	\caption{Time domain plot for data collected from inside sensor before and after inserting Red Palm Weevils through hours}
	\label{fig:r_inside}
\end{figure*}

Time series for outside and inside sensors are shown in Figures \ref{fig:r_outside} and \ref{fig:r_inside}, where y-axis is magnitude value and x-axis is the index (data order in the six hours). The time domain does allow to easily observe the signature, that is why we will use signal processing and statistical techniques to analyze the collected data. 

\begin{figure*}[ht]
	\centering
	\includegraphics[width=15cm]{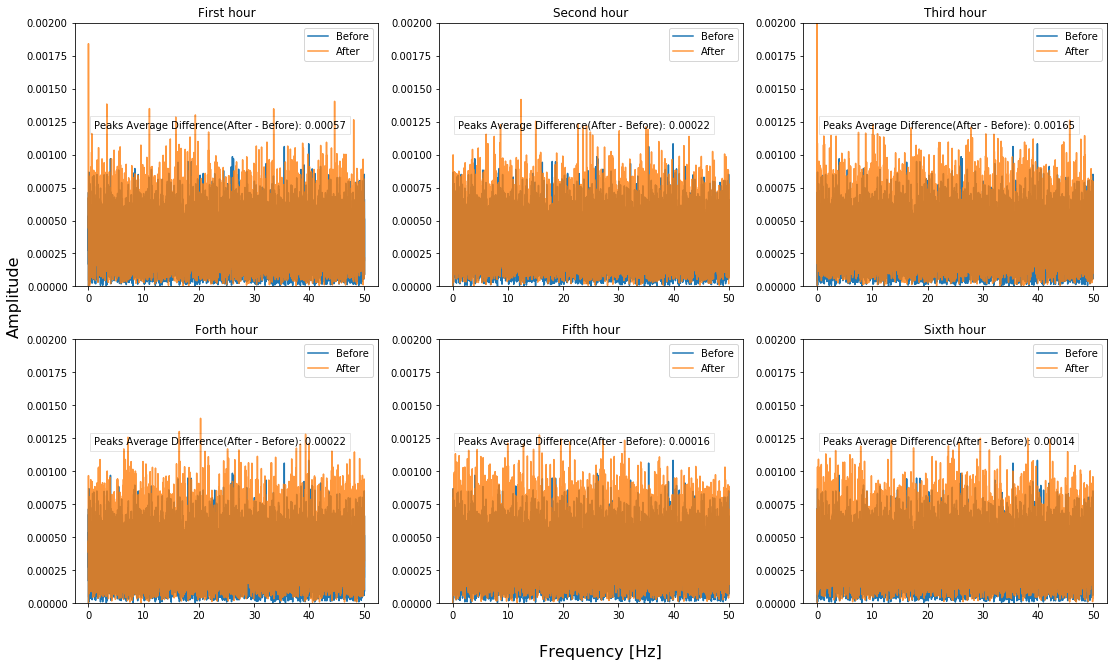} 
	\caption{FFT Frequency domain plot for data collected from outside sensor before and after inserting Red Palm Weevils through hours}
	\label{fig:fft_outside}
\end{figure*}

\begin{figure*}[ht]
	\centering
	\includegraphics[width=15cm]{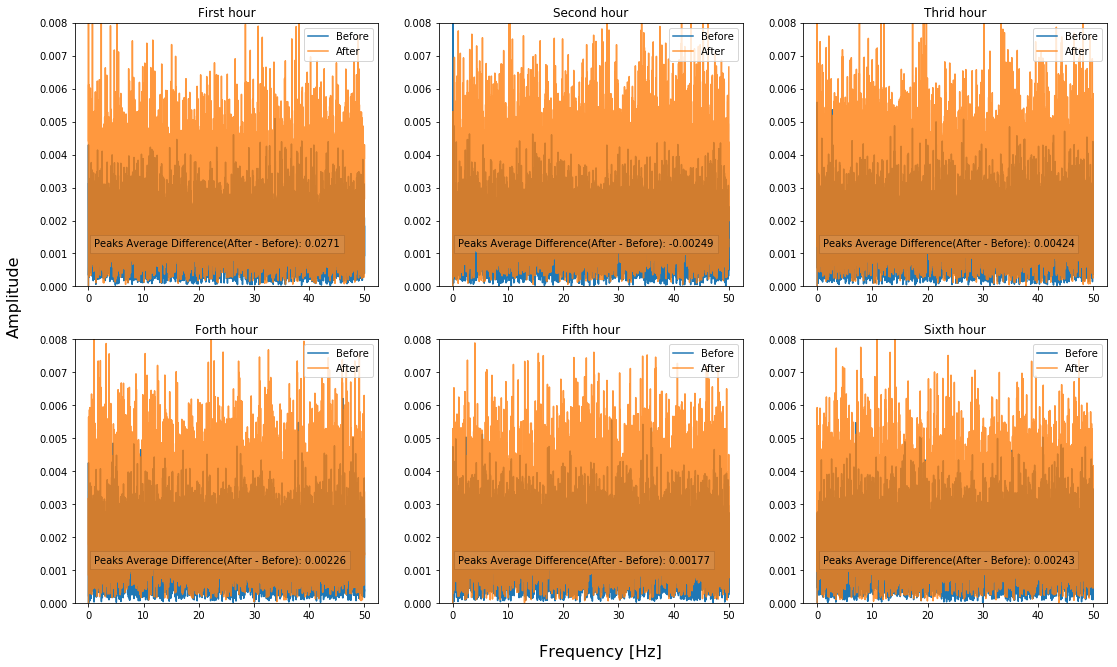} 
	\caption{FFT Frequency domain plot for data collected from inside sensor before and after inserting Red Palm Weevils through hours}
	\label{fig:fft_inside}
\end{figure*}

By applying the FFT technique, the data for the outside and the inside sensors are shown in Figures  \ref{fig:fft_outside} and \ref{fig:fft_inside}, respectively along with peaks average difference (between after, and before infestation). We can observe that when the sensor is placed inside the tree, the signature of the insect is more observable in the FFT plots as compared to an outside sensor placement. In the case of the inside sensor, it is clear that most of the FFT values remain lower than 0.004 before infestation, whereas they become higher than this threshold after infestation. 

\begin{figure*}[ht]
	\centering
	\includegraphics[width=15cm]{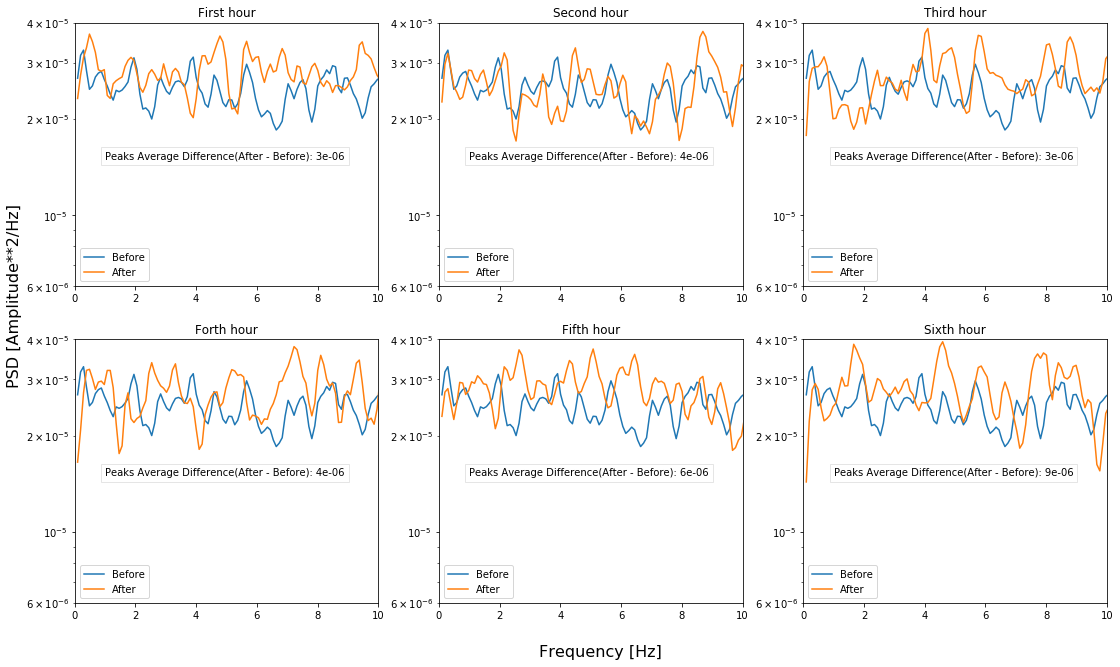} 
	\caption{PSD Frequency domain plot for data collected from outside sensor before and after inserting Red Palm Weevils through hours}
	\label{fig:psd_outside}
\end{figure*}

\begin{figure*}[ht]
	\centering
	\includegraphics[width=15cm]{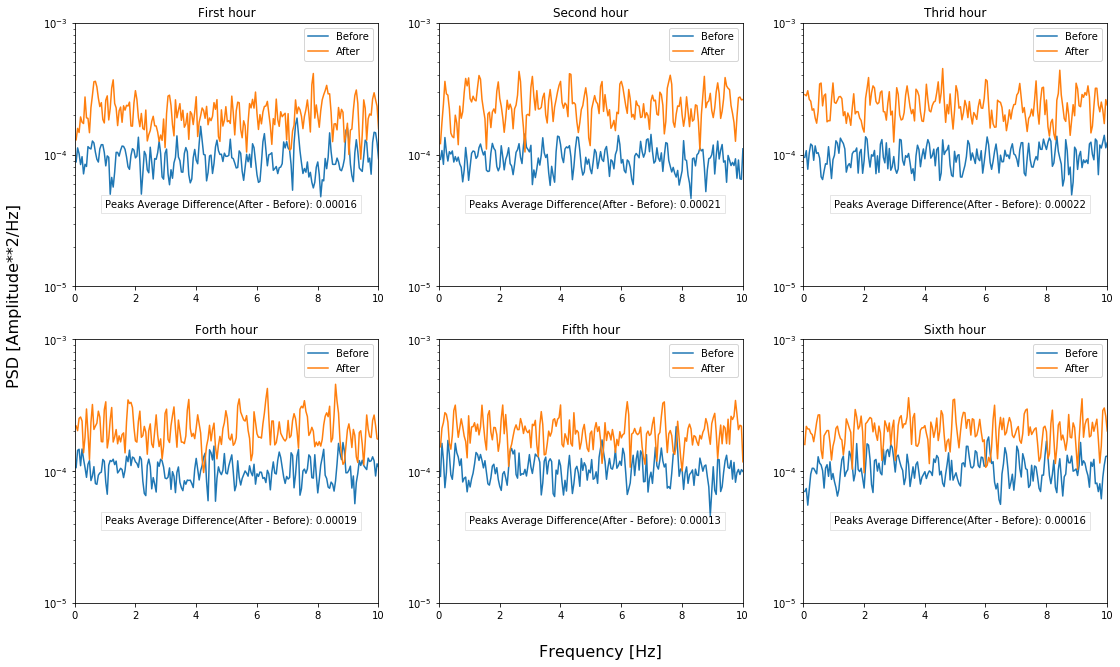} 
	\caption{PSD Frequency domain plot for data collected from inside sensor before and after inserting Red Palm Weevils through hours}
	\label{fig:psd_inside}
\end{figure*}

By applying the PSD technique, Figures \ref{fig:psd_outside} and \ref{fig:psd_inside} show the data for outside, and inside sensors, respectively, through hours along with peaks average difference (between after, and before infestation) (PAD) for the whole data values. The first 10 frequencies (0Hz - 10Hz) were plotted since RPWs movements are difficult to detect at high frequencies (more than 10Hz). First , for the outside sensor figure, we can find that PSD has very small values as compared to those of the inside sensor data. Also, the outside sensor mean peaks differences fluctuates for before and after infestation. However, the PSD of the inside sensor placement shows a clear difference of the signal between the case before infestation and the case after infestation. 

We can conclude from these two observations that outside sensor placement is more prone to external environment noise as compared to the inside sensor placement; Therefore, the inside sensor placement allows to find RPWs signature more clearly. Also, In the inside sensor placement, we can find that PSD values for data collected after infestation are much larger than data collected before infestation and the RPW signature can be seen easily in Figure \ref{fig:psd_inside}.

\subsubsection{Statistical techniques}
In this section, we present the results of the statistical analysis of the collected data. We first start by analyzing the statistical properties through box plots, then by comparing the cumulative distribution functions of the collected data before and after the insertion of red palm weevil. \textcolor{blue}{Table \ref{table3} summarizes the central tendency and the shape of distribution for all data-sets.}

\begin{table*}[]
	\begin{scriptsize}
		\centering
		\begin{tabular}{|p{1.6cm}|p{1cm}|c|p{1.1cm}|c|c|c|c|c|c|p{1.3cm}|}
			\hline
			\textbf{Data-set} & \textbf{Sample Size} & \textbf{Mean} & \textbf{Standard Deviation} & \textbf{Median} & \textbf{Minimum} & \textbf{25th percentile} & \textbf{50th percentile} & \textbf{75th percentile} & \textbf{Maximum} & \textbf{Duration (in minutes)} \\ \hline
			\textbf{Outside sensor before insertion} & 28,299 & 10.04 & 0.06 & 10.04 & 9.77 & 10.00 & 10.04 & 10.08 & 10.27 & 60 \\ \hline
			\textbf{Outside sensor after insertion} & 18,712 & 10.08 & 0.06 & 10.08 & 9.82 & 10.04 & 10.08 & 10.12 & 10.28 & 60 \\ \hline
			\textbf{Inside sensor before insertion} & 24,077 & 9.74 & 0.25 & 9.73 & 8.29 & 9.58 & 9.73 & 9.89 & 11.67 & 60 \\ \hline
			\textbf{Inside sensor after insertion} & 17,614 & 9.94 & 0.37 & 9.93 & 8.20 & 9.71 & 9.93 & 10.15 & 12.64 & 60 \\ \hline
		\end{tabular}
		\caption{Measures Of Central Tendency}
		\label{table3}
	\end{scriptsize}
\end{table*}

\begin{figure}[htb!]
	\centering
	\includegraphics[width=9cm]{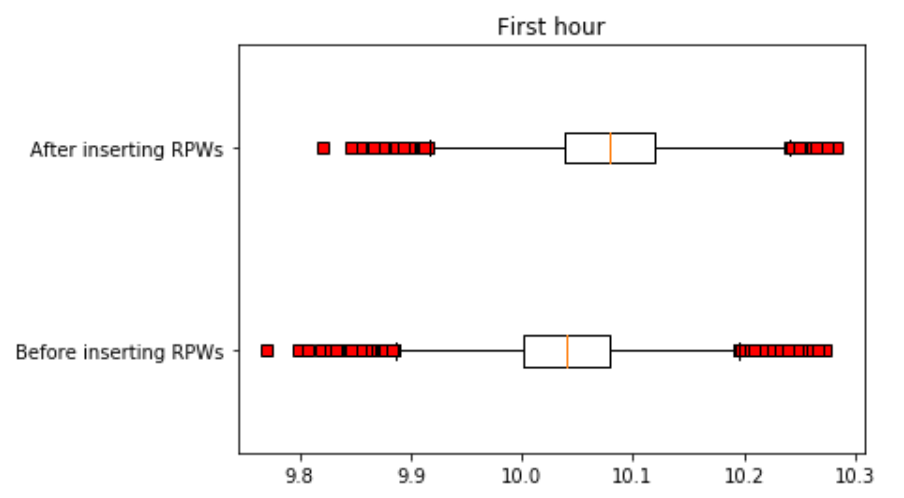} 
	\caption{Box-plot graph for the outside sensor, before and after infestation for the first hour}
	\label{fig:box-plot_outside_sensor}
\end{figure}

\begin{figure}[htb!]
	\centering
	\includegraphics[width=9cm]{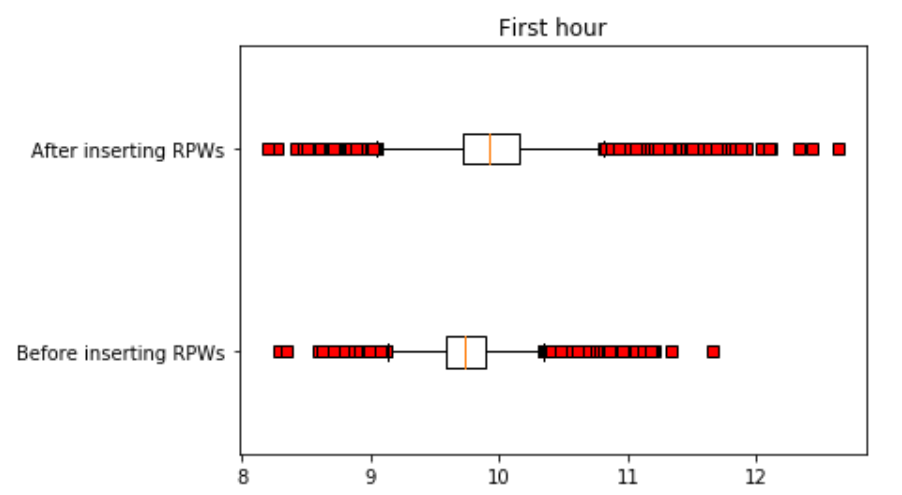} 
	\caption{Box-plot graph for the inside sensor, before and after infestation for the first hour}
	\label{fig:box-plot_inside_sensor}
\end{figure}

\paragraph{\textbf{Statistical Measures}}
We collected data before and after infestation using outside and inside sensors. Figures \ref{fig:box-plot_outside_sensor} and \ref{fig:box-plot_inside_sensor} present the box-plots corresponding to the collected data for outside and inside sensors for one hour respectively. 

We can observe that for any sensor placement, the mean and the variance of the infestation use case is larger than the case of non infestation. This is due to a more signal variation of the time series due to the acceleration induced by the motion of the insect. 

In the case before infestation, both the mean and the median are equal at 10.04, which indicates that the distribution is symmetric. However, in the case after infestation, we can notice how the outliers have more variety in their values. In addition, they are much similar in case of equality between the median and the mean, with a 0.04 increase, to become 10.08.

On the other hand, the data collected from the inside sensor (See figure \ref{fig:box-plot_inside_sensor}) has noticeable results. The subtraction between the maximum value (Third Quartile value + 1.5*IQR) and the minimum value (First Quartile value - 1.5*IQR) for the case before infestation is 1.2169, but for the case after infestation has increased to 1.7720. After all, relying on the difference between maximum and minimum seems like a promising conceivable result.

\begin{figure}[htb!]
	\centering
	\includegraphics[width=9cm]{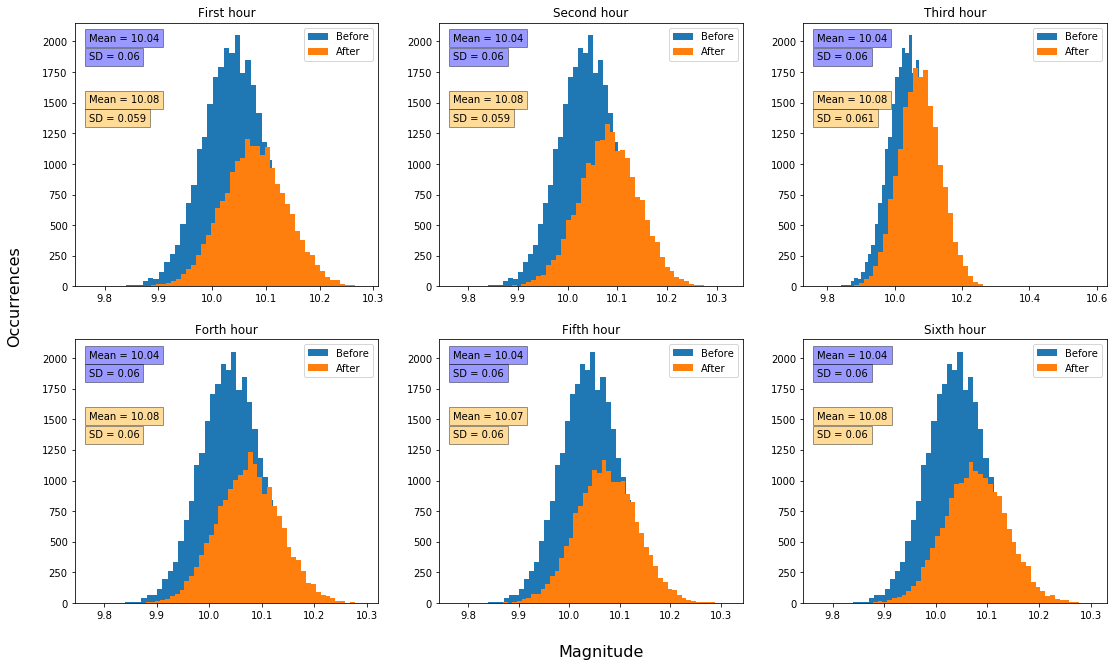} 
	\caption{Histogram representation for the outside sensor, before and after inserting the RPWs for the first six hours}
	\label{fig:Histogram_outside}
\end{figure}

\begin{figure}[htb!]
	\centering
	\includegraphics[width=9cm]{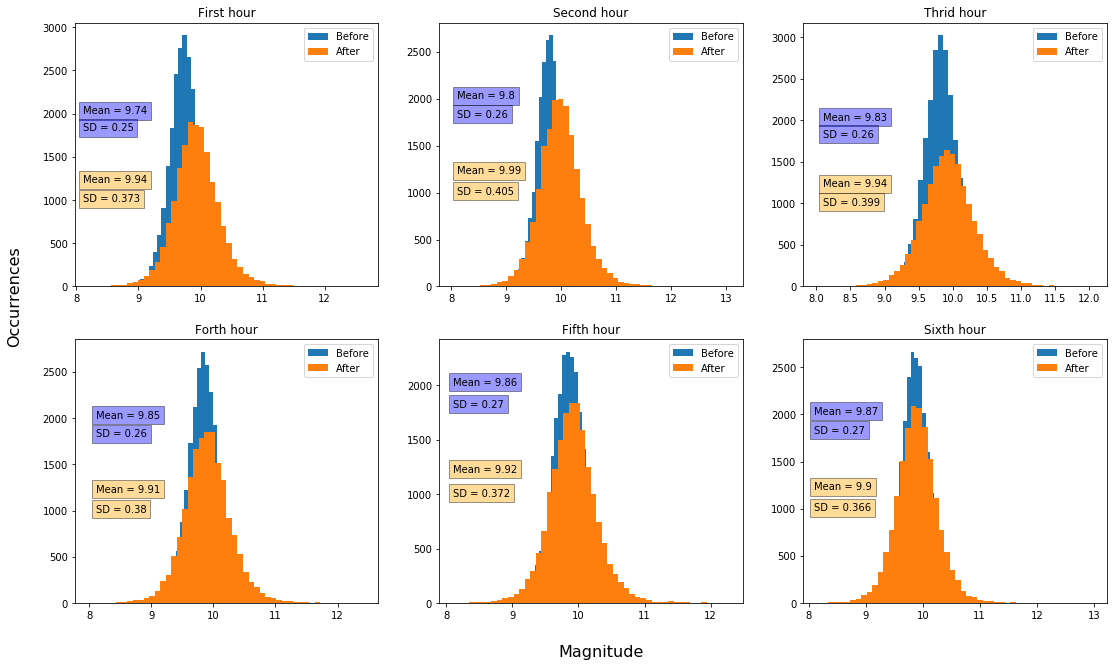} 
	\caption{Histogram representation for the inside sensor, before and after inserting the RPWs for the first six hours}
	\label{fig:Histogram_inside}
\end{figure}

\paragraph{\textbf{Probability Distributions}}
\textcolor{blue}{ Figures \ref{fig:Histogram_outside} and \ref{fig:Histogram_inside} demonstrate the Histogram graphs for inside sensor and outside sensor for six consecutive hours. We experimented the datasets using a different number of bins. We found out that 50 bins are more suitable to use.}

It is clear that the distribution of data follows a Gaussian distribution for both the inside and the outside sensor placements. However, we can observe that after infestation, the bell-shape of the normal distribution becomes more spread as compared to before infestation case. This is reasonable as data has more variation in case of infestation due to the insect motion and thus the variance is larger. This is even confirmed in the results of Table \ref{table3}. Also, we can observe that the peak values are shifted to the right in the case of infestation as the mean is larger. 

The results is consists for all the observation windows in the six plots, where each plot represents one hour of time window as mentioned. 



\paragraph{\textbf{Cumulative Distribution Functions}}

\begin{figure}[htb!]
	\centering
	\includegraphics[width=9cm]{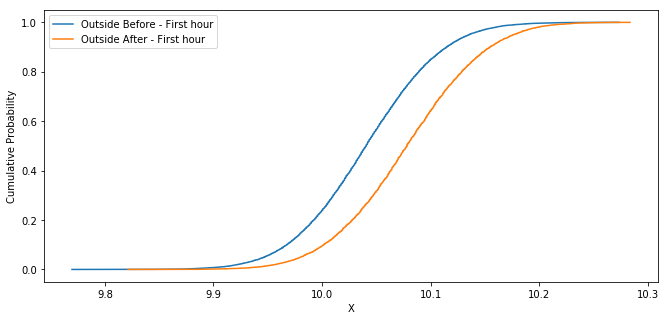} 
	\caption{Cumulative Distribution Function for the outside sensor, before and after inserting the RPWs for the first hour}
	\label{fig:CDF_outside_sensor}
\end{figure}

\begin{figure}[htb!]
	\centering
	\includegraphics[width=9cm]{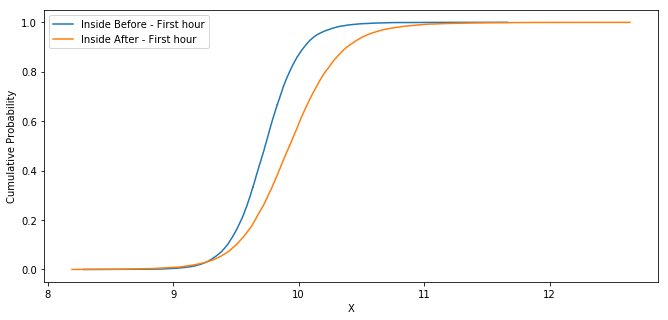} 
	\caption{Cumulative Distribution Function for the inside sensor, before and after inserting the RPWs for the first hour}
	\label{fig:CDF_inside_sensor}
\end{figure}

\textcolor{blue}{ Figures \ref{fig:CDF_outside_sensor} and \ref{fig:CDF_inside_sensor} present the Cumulative Distribution Function (CDF) for both outside and inside sensors only for the first hour. The results confirm those of the probability distribution since the CDF for before infestation and after infestation are different. The after infestation CDFs is lower that the before infestation CDF. This is because the data has more variance after infestation.}

\subsection{Lessons learned}
At the end, we conclude that the accelerometer sensor is effective in the early detection of the red palm weevil infestation in palm trees. Our initial results confirm the possibility to find clear signature of infestation using both signal processing and statistical techniques. The results of the inside sensor placement seem to be more robust that those of the outside placement, very likely due to more immunity against external noises. 
It is recommended to develop a conic metallic material that can be inserted well inside the palm tree and gets attached to an accelerometer and vibration sensor to be able to improve the features of the infestation signals. 
In addition, the IoT system that is developed in the context of this project provides a full-stack prototype for the real-time monitoring of palms trees, which combined with signal processing results of the infestation contributes uniquely to early detection of infestations. 


\section{Conclusion}

\section*{Acknowledgments}
This work is supported by the Robotics and Internet of Things Lab of Prince \cite{dawood1} Sultan \cite{dawood2-abdullah2009biological} University.

\bibliographystyle{ieeetr}
\bibliography{biblio}

\end{document}